\documentclass[structabstract]{aa}
\usepackage{txfonts}
\usepackage{graphicx}
\usepackage{amsbsy}
\usepackage{bm}
\usepackage{xcolor}
\usepackage{soul}
\DeclareMathOperator{\Cov}{Cov}

\def\mathbi#1{\textbf{\em #1}}

\usepackage{natbib}
\bibpunct{(}{)}{;}{a}{}{,}

\begin{document}

\title{AMICO galaxy clusters in KiDS-DR3: The impact of estimator statistics on the luminosity-mass scaling relation}
\titlerunning{AMICO galaxy clusters in KiDS-DR3: luminosity-mass scaling relation}
\author{Merijn Smit \inst{1}\thanks{\email{msmit@strw.leidenuniv.nl}} \and Andrej Dvornik \inst{2,1} \and Mario Radovich \inst{3} \and Konrad Kuijken \inst{1} \and Matteo Maturi \inst{4} \and Lauro Moscardini \inst{5,6,7} \and Mauro Sereno \inst{6,7}}
\institute{Leiden Observatory, Leiden University, PO Box 9513, 2300RA Leiden, The Netherlands \and Ruhr University Bochum, Faculty of Physics and Astronomy, Astronomical Institute (AIRUB), German Centre for Cosmological Lensing, 44780 Bochum, Germany \and INAF - Osservatorio Astronomico di Padova, vicolo dell’Osservatorio 5, Padova 35122, Italy \and Zentrum f\"ur Astronomie, Universit\"at Heidelberg, Philosophenweg 12, D-69120 Heidelberg, Germany \and Dipartimento di Fisica e Astronomia ``Augusto Righi'' - Alma Mater Studiorum Universit\`{a} di Bologna, via Piero Gobetti 93/2, I-40129 Bologna, Italy \and INAF - Osservatorio di Astrofisica e Scienza dello Spazio di Bologna, via Gobetti Piero 93/3, I-40129 Bologna, Italy \and INFN - Sezione di Bologna, viale Berti Pichat 6/2, I-40127 Bologna, Italy}
%Mauro Sereno \orcid{0000-0003-0302-0325}

\date{Received <date> / Accepted <date>}

\abstract
{As modern-day precision cosmology aims for statistical uncertainties of the percent level or lower, it becomes increasingly important to reconsider estimator assumptions at each step of the process, along with their consequences on the statistical variability of the scientific results.}
{We compare $L^1$ regression statistics to the weighted mean, the canonical $L^2$ method based on Gaussian assumptions, to infer the weak gravitational shear signal from a catalog of background ellipticity measurements around a sample of clusters, which has been a standard step in the processes of many recent analyses.}
{We use the shape measurements of background sources around $6925$ AMICO clusters detected in the KiDS third data release. We investigate the robustness of our results and the dependence of uncertainties on the signal-to-noise ratios of the background source detections. Using a halo model approach, we derive lensing masses from the estimated excess surface density profiles.}
{The highly significant shear signal allows us to study the scaling relation between the $r$-band cluster luminosity, $L_{200}$, and the derived lensing mass, $M_{200}$. We show the results of the scaling relations derived in 13 bins in $L_{200}$, with a tightly constrained power-law slope of $\sim 1.24\pm 0.08$. We observe a small, but significant, relative bias of a few percent in the recovered excess surface density profiles between the two regression methods, which translates to a $1\sigma$ difference in $M_{200}$. The efficiency of $L^1$ is at least that of the weighted mean and increases with higher signal-to-noise shape measurements.}
{Our results indicate the relevance of optimizing the estimator for inferring the gravitational shear from a distribution of background ellipticities. The interpretation of measured relative biases can be gauged by deeper observations, and the increased computation times remain feasible.}


\maketitle

\nocite{WLgen_BM_2017,MAD_Falk_1997}

\section{Introduction} \label{sec:intro}

% General

Statistics is an essential part of astronomy \citep{Stat_Heck_1985,Stat_Feigelson_1988,Stat_Feigelson_2009,Stat_Feigelson_2013}. The field relies on inferring physical properties, which cannot be determined directly, from observable quantities, which in turn need to be corrected for systematic effects as well as instrumental and observational biases. The key question that always needs to be answered when interpreting observations and results -- before discussing how accurately these results can be constrained -- is what one is actually seeing.

% WL general

Weak gravitational lensing, caused by the deflection of light rays by density variations along the traveled path, has been a case in point for the last three decades. Gravitational lensing is a convex focusing effect that can magnify and shear affected background sources. The observed shapes and number counts can conversely yield information about these density variations but need to be disentangled statistically from the unknown intrinsic properties of background sources, such as distance, size (and luminosity), and shape. 

The first detections of coherent alignments of galaxy shapes were observed in the background of clusters \citep{WLhist_Tyson_1990}, and subsequently in the emerging fields of galaxy-galaxy lensing (where the lensing ``structure'' is itself an ensemble of lenses; \citealt{WLhist_Brainerd_1996}) and cosmic shear (the weak lensing induced by large-scale structure; \citealt{WLhist_Wittman_2000,WLhist_Bacon_2000,WLhist_Kaiser_2000,WLhist_Waerbeke_2000}). Since then, techniques have progressed rapidly, and demands on accuracy have become increasingly stringent.

% Introduce subtopic

This is the second in a set of papers wherein we focus on the statistical aspects of inferring the lensing signal from the intrinsic shapes and the estimated lensing geometry, which depends on the distances between the observer, the moment of deflection, and the background sources. Assuming the cosmological principle, the intrinsic shapes of a sample\footnote{There are several considerations involved in the proper selection of such a sample, as explained in Sects. \ref{sec:wl_stat} and \ref{sec:data}.} of background galaxies, including their orientation, are random, and the intrinsic galaxy shapes should average out from a sufficiently large sample, leaving the weak lensing signal as a net ellipticity. The common approach has been to take a weighted mean of galaxy ellipticities, which has computational and analytical advantages and, most importantly, is an unbiased estimator of the shear in the absence of pixel noise in the galaxy images \citep{WLgen_SS_1997}.

In practice, however, there are many sources of noise and the mean is known to be biased, underestimating the underlying shear signal \citep{Bias_Melchior_2012,Bias_Viola_2014,Bias_Sellentin_2018,Bias_Mandelbaum_2018}. The distribution of intrinsic galaxy shapes is well known to be non-Gaussian \citep{Shape_Lambas_1992,Shape_Rodriguez_2013} and, in fact, centrally peaked. In \citet[][hereafter Paper I]{PaperI}, we explored alternative estimators besides the mean that could potentially be better suited for such a cuspy distribution. It was found, using realistic simulated distributions and resampling of Canada-France-Hawaii Lensing Survey (CFHTLenS) shape measurements \citep{CFHTLenS_Heymans_2012}, that $L^1$ norm regression, also known as least absolute deviations (LAD), reduces bias from between $\sim -4\%$ and $\sim -4.5\%$ to between $\sim +1\%$ and $\sim -3\%$, while at the same time reducing uncertainty by $\sim 9\%$ to $\sim 23\%$.

%\LEt{The official acronym, even though referring to plural, is LAD.}

% Goal of this research

In this paper we extend this study by applying these statistics to a weak lensing analysis of 6925 galaxy clusters in the Adaptive Matched Identifier of Clustered Objects (AMICO) cluster catalog \citep{AMICO_Bellagamba_2011,AMICO_Radovich_2017,AMICO_Bellagamba_2018,AMICO_Maturi_2019,AMICO_Bellagamba_2019} of the third data release of the Kilo-Degree Survey \citep[KiDS-450;][]{KiDS_de_Jong_2017}. As opposed to Paper I, in this case the true lensing signal (here in the form of the excess surface density of the clusters) is unknown. We therefore study the relative biases and uncertainties between LAD and the mean, and we compare results to our findings in Paper I.

An important application is then to study the relation between the observable properties of clusters and groups and the physical quantities derived from the lensing signal (i.e., the matter distribution) to better our understanding of galaxy and cluster formation and cosmological models \citep[e.g.,][]{Form_Kautsch_2008,Form_Leauthaud_2010,AMICO_Lesci_2020}. We calculate halo masses from the obtained lensing signals and derive a scaling relation between the observed $r$-band luminosity and the lensing mass, investigating the impact of estimator choice on the resulting constraints.

% Zoom out: how does this bias compare to other steps in the process?

The order of magnitude of this estimated bias in the weak lensing results can be dominant compared to other sources of uncertainty in the process. Developments in the field have led to current constraints of the multiplicative bias in shape measurements  on the order of $\sim 1\%$ \citep{Shape_Bernstein_2002,Shape_Hirata_2003,STEP_Heymans_2006,STEP_Massey_2007,lensfit_Miller_2007,lensfit_Kitching_2008,GREAT_Bridle_2010,SYST_Voigt_2010,SYST_Bernstein_2010,GREAT_Kitching_2012,Bias_Kacprzak_2012,Bias_Melchior_2012,Bias_Refregier_2012,SYST_Heymans_2012,GREAT_Mandelbaum_2015,KiDS_Viola_2015,KiDS_Fenech_Conti_2017}. The uncertainty in the lensing geometry between the observer, lens, and background sources, introduced by the estimation of the photometric redshift probability distributions, can be a few percent \citep[][and Appendix \ref{app:photoz}]{KiDS_Hildebrandt_2017,AMICO_Bellagamba_2019}. The broad category of selection biases, for example those introduced by intrinsic alignments, contamination of the background sample by cluster member galaxies, blending, detection, and subsequent selection effects, typically accumulate up to a few percent \citep{Miyatake_2015,KiDS_Uitert_2017,AMICO_Bellagamba_2019} for cluster weak lensing. For instance, estimations on background selection yield a foreground contamination on the order of $2\%$, which can be partly corrected for, but does increase the uncertainty \citep[][and Appendix \ref{app:clus_bg}]{KiDS_Dvornik_2017,AMICO_Bellagamba_2019}. In this study we investigate the usability of background sources to radii smaller than in \citet{AMICO_Bellagamba_2019}.

These demands on accuracy and precision become higher as the data yield, and therefore the statistical power of surveys, increases dramatically \citep{Bias_Mandelbaum_2018}, as achieved by COSMOS\footnote{http://cosmos.astro.caltech.edu/} \citep{COSMOS_2007}, CFHTLenS\footnote{http://www.cfhtlens.org} \citep{CFHTLenS_Heymans_2012}, RCSLenS\footnote{http://www.rcslens.org/} \citep{RCSLenS_2016}, KiDS\footnote{http://kids.strw.leidenuniv.nl/} \citep{KiDS_de_Jong_2013}, and DES\footnote{http://www.darkenergysurvey.org/} \citep{DES_2016}, and foreseen for future surveys such as LSST\footnote{https://www.lsst.org/} \citep{LSST_2019} and Euclid\footnote{http://www.euclid-ec.org/} \citep{EUCLID_2011}. While these two future surveys will require constraints on systematic uncertainty of order $\leq 2\times 10^{-3}$ \citep{Bias_Mandelbaum_2018}, we show that, even for weak lensing analyses in the last decade, the bias in shear inference can dominate other sources, such as the aforementioned multiplicative shape measurement bias that is commonly corrected for, as in \citet{KiDS_Viola_2015}, \citet{KiDS_Dvornik_2017}, and \citet{AMICO_Bellagamba_2019}.

% Other approaches and importance of the perspective from this research

Several other approaches have been made to address this, including analytic modeling of the bias \citep[e.g.,][]{Bias_Viola_2014}, weight corrections and priors \citep{WLhist_Bonnet_1995,WLhist_Waerbeke_2000,Shape_Bernstein_2002}, or nulling techniques \citep{Shape_Herbonnet_2017}. The calculation of the main observable, the shapes of lensed background sources, itself relies on statistical methods. These are based mainly on surface brightness moments \citep{Shape_Kaiser_1995, Shape_Rhodes_2000} or model fitting \citep{Shape_Kuijken_1999,Shape_Bernstein_2002,Shape_Hirata_2003,Shape_Refregier_2003,Shape_Kuijken_2006,lensfit_Miller_2007,lensfit_Kitching_2008}. This means the most common approaches are corrections on a statistic that remains fundamentally skewed \citep{Bias_Sellentin_2018,Bias_Mandelbaum_2018}.

Promising alternative approaches by \citet{Shape_Bernstein_2014} and \citet{Shape_Schneider_2015} do not reproduce individual background shapes, but directly determine the underlying shear field from ensembles of background sources, reconsidering these steps in the chain of statistical inference. While future lensing surveys will require innovative improvements, these methods and their priors need to be gauged by deep observations of high signal-to-noise, and it is of fundamental importance that these calibrations are well constrained and do not suffer from even subtle systematic biases. In other words, the comparison of several perspectives is paramount in determining what we actually see.

% PAPER ORG

The remainder of this paper is organized as follows. We introduce the definitions of galaxy shapes and the weak lensing formalism in Sect. 2 and relate these to our statistical approach. Data, analysis methods, and selection criteria are described in Sect. 3, while Sect. 4 states our results and analysis. Section 5 gives a summary of our conclusions.

Throughout this paper we assume a Planck \citep{Planck_XVI_2014} cosmology with $\Omega_{\mathrm{M}}=0.315$, $\Omega_{\mathrm{{\Lambda}}}=0.685$, and $H_0=100.0$ $h$ km s$^{-1}$ Mpc$^{-1}$. All measurements are in comoving units, unless specifically noted otherwise, such as in Sect. \ref{sec:wl_stat}.

% Note that in the literature, I see mostly comoving, and only a few times comoving. Even on ADS in A&A articles, I see both versions appearing. I would suggest to keep with the most common occurence, but I have no principal objections.

\section{Weak gravitational lensing statistics} \label{sec:wl_stat}

We briefly review the principles of weak gravitational lensing and relate the central concepts to our statistical approach, introducing the terminology and notation conventions used in this paper. We refer the reader to excellent reviews, such as \citet{WLgen_BS_2001}, \citet{WLgen_S_2006}, \citet{WLgen_HJ_2008}, and \citet{WLgen_BM_2017}, for more in-depth approaches.

\subsection{Principles of weak lensing} \label{sec:wl_gen}

\begin{figure}[h]
\centering
\resizebox{\hsize}{!}{
\includegraphics{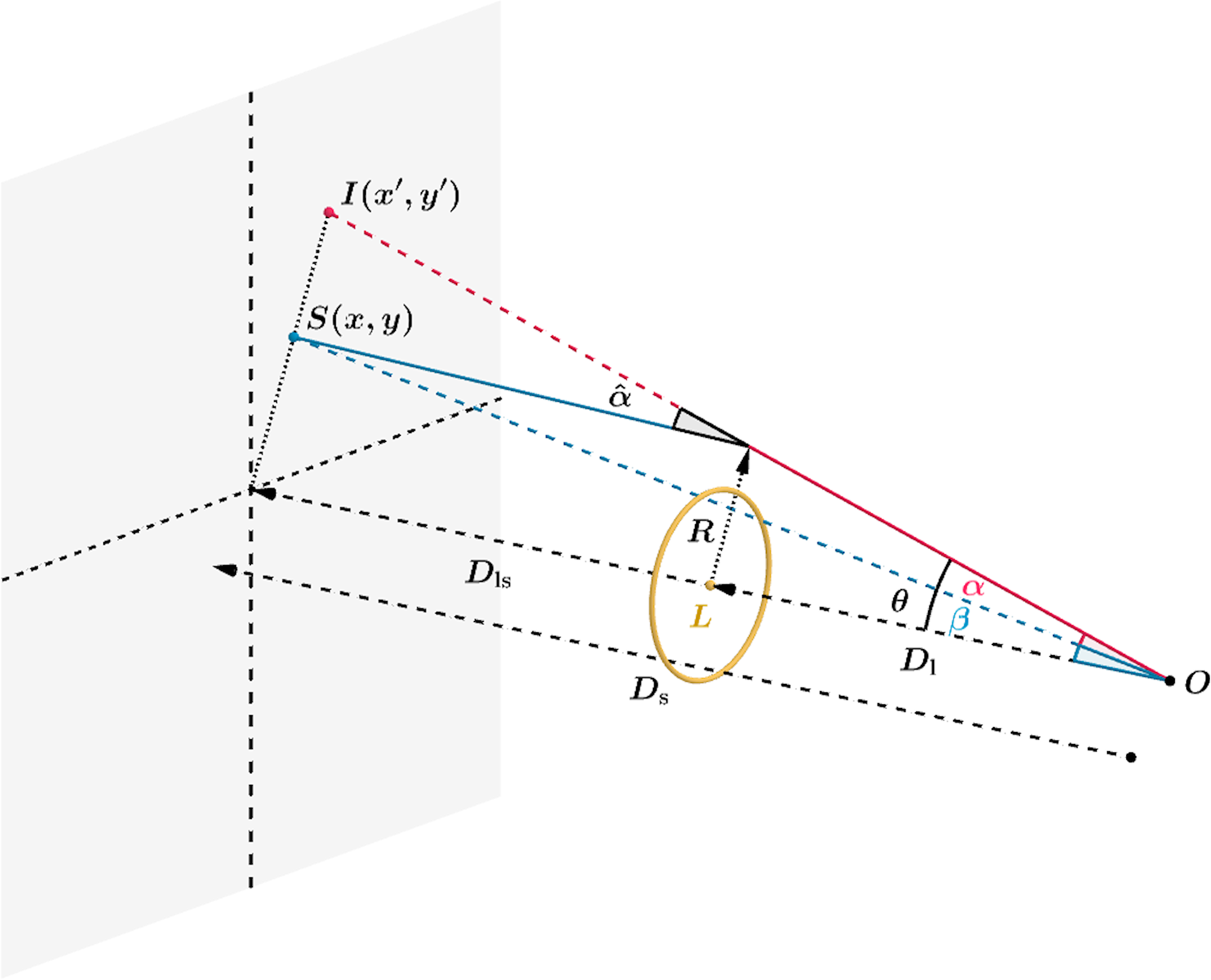}
}
\caption{Representation of a gravitational lens system, showing the displacement of a source at position $S(x,y)$ to an image at position $I(x',y')$, where we take the origin of the source plane to be collinear with the position of the lens, $L$, and the observer, $O$.}
\label{fig:lens_geo}
\end{figure}

Rays of light are deflected by the curvature or space-time due to mass inhomogeneities along their path. A mass overdensity acts as a convex lens on the light rays from distant sources behind that lens to an observer. In this section, we use $D_{\mathrm{l}}$  to denote angular-diameter distances from the observer to the lens, $D_{\mathrm{ls}}$ from the lens to the background source, and $D_{\mathrm{s}}$ from the observer to the background source (see Fig. \ref{fig:lens_geo}), and in the remainder of this paper we translate quantities to comoving units where necessary.

For the purposes of this work, the extent of the lensing mass along the line of sight, compared to the distances from the observer to the lens and from the lens to the background source, can be considered negligible. In this so-called thin-lens approximation, the deflection of light rays by a deflection angle, $\hat{\vec{\alpha}}$, leads to an effective angular displacement (again, see Fig. \ref{fig:lens_geo}), \begin{equation}
  \vec{\alpha} = -\frac{D_{\mathrm{ls}}}{D_{\mathrm{s}}} \hat{\vec{\alpha}}\,,
\end{equation}also called the reduced deflection angle, which relates the observed position, $\vec{\theta}$, of a distant point source to its unlensed position, $\vec{\beta}$, by the lens equation \begin{equation}
  \label{eq:lens}
  \vec{\beta}=\vec{\theta}-\vec{\alpha}.
\end{equation}

It can be shown through the relation between $\hat{\vec{\alpha}}$ and the three-dimensional gravitational potential, $\Phi$, that this displacement is then described by $\vec{\alpha}=\vec{\nabla}_{\theta}\, \psi$, where \begin{equation}
  \label{eq:lens_pot}
  \psi = \frac{2}{c^2}\frac{D_{\mathrm{ls}}}{D_{\mathrm{l}}D_{\mathrm{s}}}\int \Phi\, dz
\end{equation}is called the (two-dimensional) lensing potential.

The differential effect of the deflection of light on the images, $I(x,y)$, of extended background sources can to first order be described as a coordinate transformation by taking the derivatives in the lens equation (Eq. \ref{eq:lens}) of the original angular position, $\beta $, with respect to the observed position, $\theta $. Substituting $\vec{\nabla}_{\theta}\, \psi$ for $\vec{\alpha}$, we obtain the Jacobian matrix,\begin{equation}
  \label{eq:coor_tr}
  \left( \begin{array}{c} x' \\ y' \end{array} \right) =\left( \begin{array}{cc} 1-\psi_{11} & -\psi_{12} \\ -\psi_{21} & 1-\psi_{22} \end{array} \right) \left( \begin{array}{c} x \\ y \end{array} \right) \,,
\end{equation} with\begin{equation}
  \psi_{ij}=\frac{\partial^2\psi}{\partial\theta_i\partial\theta_j} \,,
\end{equation} resulting in the lensed image $I(x',y')$, which is the key observable in our weak lensing study.

\subsubsection{Critical surface mass density} \label{sec:wl_sigcr}

To interpret the effect on the source image, we note that such a transformation can be decomposed into three parts, namely the identity ($\mathbi{I}$), an isotropic part that describes a multiplication, and an anisotropic traceless part that describes a shearing of the image:\begin{equation}
  \label{eq:iso_aniso}
\mathbi{I} - \frac{1}{2}(\psi_{11}+\psi_{22})\mathbi{I} + \left( \begin{array}{cc} -\frac{1}{2}(\psi_{11}-\psi_{22}) & -\psi_{12} \\ -\psi_{21} & \frac{1}{2}(\psi_{11}-\psi_{22}) \end{array} \right)
.\end{equation}
% In answer to your question, started should be start in the next sentence, as it is a `general method':

To relate $\psi_{ij}$ with the density of the lensing mass, we start with the isotropic term, which is half the Laplacian of the lensing potential: $\frac{1}{2}(\psi_{11}+\psi_{22})=\frac{1}{2}\nabla_{\theta}^2 \, \psi $. From Eq. \ref{eq:lens_pot}, we obtain \begin{equation}
  \frac{1}{2}\nabla_{\theta}^2 \, \psi = \frac{1}{c^2}\frac{D_{\mathrm{l}}D_{\mathrm{ls}}}{D_{\mathrm{s}}}\int 4\pi G\rho\, dz \,,
\end{equation} which is a dimensionless quantity. Defining the surface mass density as \begin{equation}
  \Sigma \equiv \int \rho\, dz
\end{equation} and gathering the rest of the right-hand side into \begin{equation}
  \frac{4\pi G}{c^2}\frac{D_{\mathrm{l}}D_{\mathrm{ls}}}{D_{\mathrm{s}}} \equiv \Sigma_{\mathrm{cr}}^{-1} \,,
\end{equation} with $\Sigma_{\mathrm{cr}}$ being the critical surface mass density, we find that the isotropic term can be written as \begin{equation}
  \label{kappa}
  \kappa \equiv \frac{1}{2}\nabla_{\theta}^2 \, \psi = \frac{\Sigma}{\Sigma_{\mathrm{cr}}} \,,
\end{equation} with $\kappa$ a normalized dimensionless surface mass density. Recognizing that $\nabla_{\theta}^2 \, \psi = \vec{\nabla} \cdot \vec{\alpha}$ is the divergence of the deflection of the light rays (i.e., the manner in which those light rays converge due to the lensing effect), $\kappa $ is simply the convergence.

\subsubsection{Shear and intrinsic ellipticity} \label{sec:wl_intr}

The shear matrix in Eq. \ref{eq:iso_aniso} has two independent components, simply called the shear $\gamma = \gamma_1+i\gamma_2$, with $\gamma_1 = \frac{1}{2}(\psi_{11}-\psi_{22})$ and $\gamma_2 = \psi_{12} = \psi_{21}$. Equation \ref{eq:coor_tr} then becomes \begin{equation}
  \left( \begin{array}{c} x' \\ y' \end{array} \right) =\left( \begin{array}{cc} 1-\kappa-\gamma_1 & -\gamma_2 \\ -\gamma_2 & 1-\kappa+\gamma_1 \end{array} \right) \left( \begin{array}{c} x \\ y \end{array} \right) .\end{equation}

This transformation leads to the magnification and distortion of the light distribution of background sources. In this work, we focus on the most commonly used net distortion or reduced shear $g = g_1+ig_2 \equiv (\gamma_1 +i\gamma_2) / (1 - \kappa)$, \begin{equation}
\label{eq:lens_re}
\left( \begin{array}{c} x' \\ y' \end{array} \right) = \left( 1-\kappa \right) \left( \begin{array}{cc} 1-g_1 & -g_2 \\ -g_2 & 1+g_1 \end{array} \right) \left( \begin{array}{c} x \\ y \end{array} \right) \,,
\end{equation}where the transformation is written as a multiplication of $( 1-\kappa )$ and a distortion matrix describing the alignment of lensed sources in the foreground potential.

The effect on a circular source is a shearing into an ellipse with axis ratio $q=\frac{b}{a}$ as \begin{equation}
\label{eq:axis}
q=\frac{1-|g|}{1+|g|} \quad \Leftrightarrow \quad |g|=\frac{1-q}{1+q}=\frac{a-b}{a+b} \,
\end{equation} and position angle $\varphi$ via \begin{equation}
\label{eq:posang}
g=|g|\left( \cos{2\varphi}+i\sin{2\varphi}\right) .
\end{equation}

As mentioned before, we do not measure this gravitational distortion directly. Background sources have an intrinsic shape distribution, and we effectively measure the combined effect of their intrinsic shape and a weak lensing distortion. It is adequate to describe images by their quadrupole brightness moments or their ellipticities as well as the respective response to weak shear distortions. It is straightforward to use the common definition\footnote{An alternative definition of ellipticity is often denoted as $|\chi|=\frac{1-q^2}{1+q^2}$, related to the geometrical eccentricity, and called polarization \citep[e.g.,][]{WLgen_SS_1995,Bias_Viola_2014}.} of ellipticity, defined as the reduced shear needed to create the intrinsic shape $\epsilon=\epsilon_1+i\epsilon_2$ of a source from an image with circular isophotes \citep{Shape_Bernstein_2002,Shape_Kuijken_2006}. The resulting ellipticity, $\epsilon$, after transforming an image with intrinsic\footnote{We note that our notation differs from Paper I. Here, the measured ellipticity is denoted as $\epsilon$, instead of $\tilde{e}$ (Paper I), and the intrinsic ellipticity is denoted as $\epsilon^I$, instead of $e$.} ellipticity $\epsilon^I$ by a distortion, $g$, is then given by \citep{WLgen_SS_1997}

\begin{equation}
\label{eq:SS3_97}
\epsilon = \frac{\epsilon^I+g}{1+g^*\epsilon^I} \quad \mathrm{for} \quad |g|\le 1 \,,
\end{equation}with $g^*$ the complex conjugate of $g$.

The intrinsic shape distribution is called the shape noise and, assuming no preferred direction on the sky, should average to zero: $\left< \epsilon^I\right> =0$. This way, each background shape measurement, $\epsilon$, is then an independent estimate of the underlying reduced shear, $g$.

In this paper we make use of the fact that the lensing signal is weak (i.e., $\kappa \ll 1$) and assume $g\approx \gamma$.
%\LEt{ single-sentence paragraph.}
% It is a single sentence, but it does not belong in the paragraph above it, nor in the next section. It's a similar message as the end of Sect. 1, "Throughout this paper...[etc.]"

\subsection{Estimation of the surface density profile} \label{sec:ESD}

The shear induced by gravitational lensing is sensitive to the density contrast. For an axisymmetric lens, we can write $|\gamma|(R)=\overline{\kappa}(\le R)-\kappa(R)$, where $\overline\kappa$ is the average convergence within radius $R$. In fact, this relation holds for other mass distributions if we average azimuthally around the lens. In this work, we study the stacked signal of many lenses and assume a net axisymmetry \citep[see, e.g.,][for weak lensing studies on elliptical lenses]{Halo_Evans_2009,Halo_Oguri_2010,Halo_Clampitt_2016,KiDS_Uitert_2017}.

Since it can be seen from Fig. \ref{fig:lens_geo} that the gravitational shear acts in the radial direction, we define the tangential and cross components of the shear as \begin{equation}
\label{eq:tang_shear}
\left( \begin{array}{c} \gamma_+ \\ \gamma_\times \end{array} \right) =\left( \begin{array}{rr} -\cos(2\phi) & -\sin(2\phi) \\ \sin(2\phi) & -\cos(2\phi) \end{array} \right) \left( \begin{array}{c} \gamma_1 \\ \gamma_2 \end{array} \right) \,,
\end{equation} with  $\phi$ the counterclockwise angle between the positive $x$ axis\footnote{Of the coordinate system in which $\gamma_1$ and $\gamma_2$ are defined.} and the vector from lens to source. This gives \begin{equation}
  \overline{\Sigma}(\le R) - \Sigma(R) \equiv \Delta\Sigma (R) = \gamma_+(R)\, \Sigma_{\mathrm{cr}} \,,
\end{equation}with $\Delta\Sigma (R)$ the excess surface density (ESD) at a radius $R$ around the lensing mass. In axisymmetric lenses, the cross component of the shear cannot arise from gravitational lensing and should average to zero, if only produced by intrinsic source orientations, and can therefore be used as an indication of systematic effects, such as imperfect corrections for the point-spread function \citep[PSF;][and Appendix \ref{app:cross}]{WLgen_S_2003}.

The ESD is then estimated using the observed ellipticities of an ensemble of sources around the lens \begin{equation}
  \Delta\Sigma (R) = \left< \epsilon_+\, \Sigma_{\mathrm{cr,ls}}\right>(R) \,,
\end{equation}with each $\epsilon_+\Sigma_{\mathrm{cr,ls}}$ an independent, albeit noisy, estimate of the ESD. Here, $\langle\cdot\rangle$ denotes a weighted average, with weights to be specified.

The $\Sigma_{\mathrm{cr,ls}}$ behaves as a geometric scaling factor, indicating the lensing efficiency for each lens-source combination. Since the variance of the noise in $\Delta\Sigma$ is then affected by $\Sigma_{\mathrm{cr,ls}}^{2}$ , the relative precision, or inverse variance, carried by each $\epsilon_+$ scales as $\Sigma_{\mathrm{cr,ls}}^{-2}$.

In this paper we study the ESD profile in comoving radial bins, and we therefore use the comoving critical surface density\begin{equation}
  \Sigma_{\mathrm{cr,com}}=\left(1+z_{\mathrm{l}}\right)^2\Sigma_{\mathrm{cr,prop}}.
\end{equation}In practice, the distance to each background source is not known exactly and is estimated by its redshift probability distribution, $p\left(z_{\mathrm{s}}\right)$. Taking this into account, we estimate the comoving critical surface density via\begin{equation}
\label{eq:sig_cr}
\left<\Sigma_{\mathrm{cr,ls}}^{-1}\right> =  \frac{4\pi G}{c^2} D(z_{\mathrm{l}}) \left(1+z_{\mathrm{l}}\right)^2 \int \frac{D(z_{\mathrm{l}},z_{\mathrm{s}})}{D(z_{\mathrm{s}})} p(z_{\mathrm{s}})dz_{\mathrm{s}}.
\end{equation}
% Above the equation: estimate, as it is what people do in general and in recent literature.

\begin{figure}[h]
\centering
\resizebox{\hsize}{!}{
\includegraphics{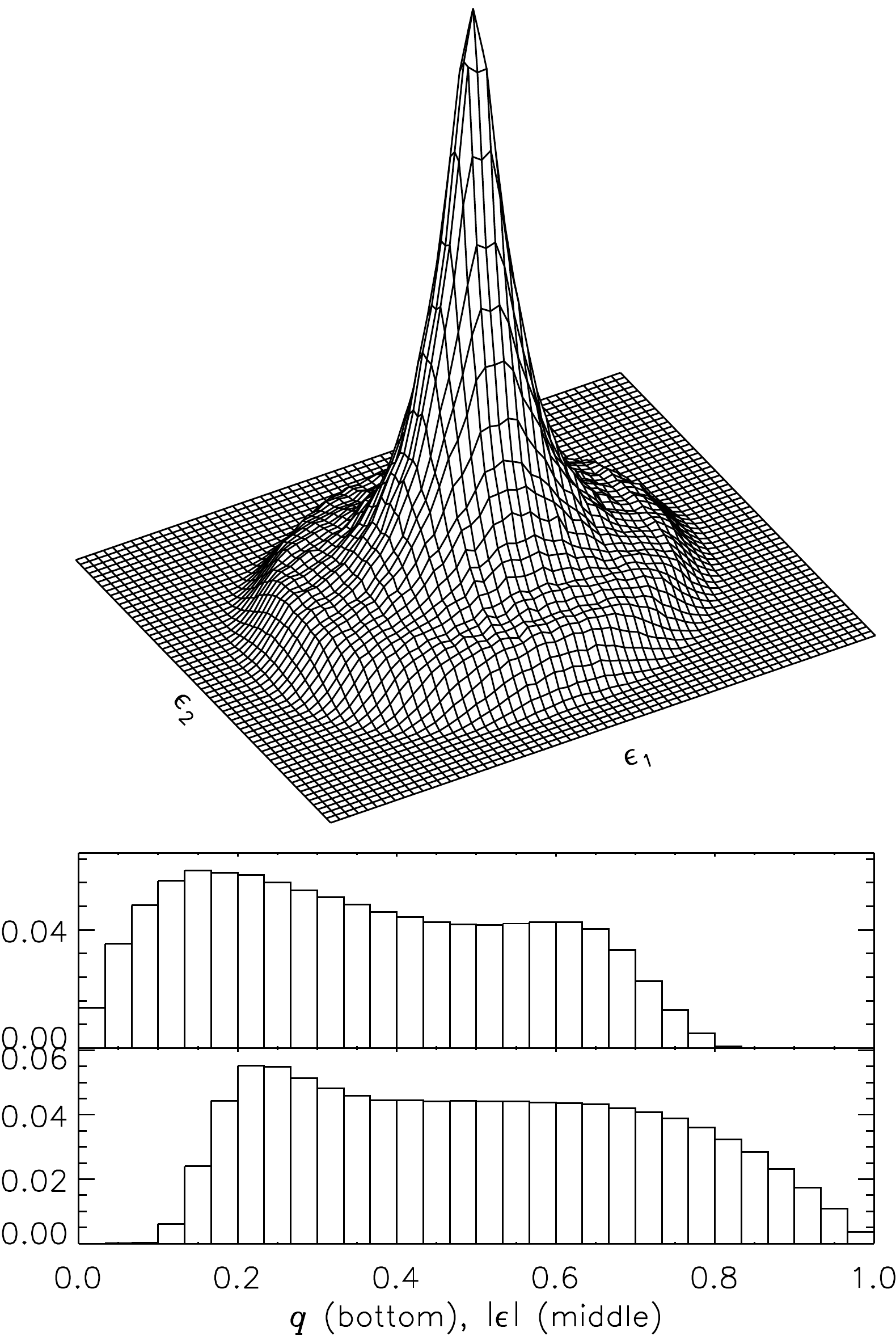}
}
\caption{Estimated ellipticity and axis ratio distributions of sources in the KiDS-450 catalog. Top: Two-dimensional histogram of ellipticities. Middle: Histogram of the absolute ellipticity, $|\epsilon|$. Bottom: Histogram of the ellipse axis ratio, $q$.}
\label{dist_kids}
\end{figure}

%BB 15 21 470 699

\subsection{Statistical framework} \label{sec:stat}

In this section we discuss the estimation of $\left< \epsilon_+\, \Sigma_{\mathrm{cr,ls}}\right>$. We refer to Paper I for a complementary discussion.

Important aspects of a good estimator, $\hat{\epsilon}$, are: (i) minimal bias, defined as the difference between the expected value of the estimator, $\left< \hat{\epsilon}\right>$, and the value of the quantity being estimated, for instance the shear ($\gamma)$ or, in this case, $\Delta\Sigma$; (ii) high efficiency, proportional to the inverse variance of the estimator, $\sigma^{-2}_{\hat{\epsilon}}$; and (iii) robustness, meaning the estimator retains these properties for a sufficient range of likely parameter distributions.

\subsubsection{Bias} \label{sec:bias}

Even though the measured ellipticity, $\epsilon$, is not a linear combination of the intrinsic shape, $\epsilon^I$, and the shear, $\gamma$, it can be shown \citep{WLgen_SS_1997} that, in the absence of further uncertainties, the mean $\mu\left(\epsilon\right)$ is an unbiased estimator for the underlying shear and that this is independent of the intrinsic ellipticity distribution, $P\left( \epsilon^I\right)$. In the canonical approach, the ESD is therefore estimated as a weighted mean of an ensemble of lens-source combinations,\begin{equation}
\label{eq:ESD_lsq}
\Delta\Sigma (R) = \frac{\sum_{\mathrm{ls}} w_{\mathrm{ls}} \epsilon_{+\mathrm{,ls}} \Sigma_{\mathrm{cr,ls}}}{\sum_{\mathrm{ls}} w_{\mathrm{ls}}} \,,
\end{equation}where we use\begin{equation}
\label{eq:w_ls}
w_{\mathrm{ls}} = w_{\mathrm{s}} \left<\Sigma_{\mathrm{cr,ls}}^{-1}\right>^2 \,.
\end{equation}Here, the weight $w_{\mathrm{s}}$ is assigned to each measured ellipticity, scaled by the estimated lensing efficiency, as explained in the previous section \citep[see, e.g.,][]{KiDS_Viola_2015,KiDS_Dvornik_2017,AMICO_Bellagamba_2019}.

In practice, there are various sources of uncertainty at each step of the process, such as source selection bias, distortion by the PSF, and biases due to the measurement pipeline. These lead to convolutions of the ellipticity distribution, before and after the gravitational lensing effect. The result is a bias in the mean as an estimate of the ESD \citep{Bias_Melchior_2012,Bias_Refregier_2012,Bias_Kacprzak_2012,Bias_Viola_2014,Bias_Kacprzak_2014}. In this case, the intrinsic shape distribution will play a role.

The weighted mean, $\mu$, is a statistic that, for a set of measurements $\epsilon_i$ with weights $w_i$, finds the estimate of $\gamma $ that minimizes the loss function \begin{equation}
\label{eq:loss_lsq}
S_{\mu}=\sum_iw_i \left[(\epsilon_{i,1}-\gamma_1)^2+(\epsilon_{i,2}-\gamma_2)^2\right] \,,
\end{equation}that is, it is a least squares (LSQ) or $L^2$ norm regression method and arises naturally as the optimal estimator for Gaussian distributions.

Figure \ref{dist_kids} shows that the measured ellipticity distribution, $P(\epsilon)$, displays crucial differences with a Gaussian distribution, showing a sharp peak and a slower decline, including a higher number of high ellipticities, $|\epsilon|$. This central peak is an unbiased tracer of the underlying shear (Paper I). By Eq. \ref{eq:loss_lsq}, the mean is sensitive to outliers and therefore not robust when inferring the shear.

In contrast, LAD or $L^1$ norm regression minimizes the loss function,\begin{equation}
\label{eq:loss_lad}
S_{\mathrm{LAD}}=\sum_iw_i \sqrt{(\epsilon_{i,1}-\gamma_1)^2+(\epsilon_{i,2}-\gamma_2)^2} \, .
\end{equation}The LAD estimate is also known as the median in one dimension or the spatial median in higher dimensions. This estimator is more sensitive to the peak and less sensitive to high ellipticity outliers. Where the mean is expected to be biased low \citep{Bias_Melchior_2012,Bias_Refregier_2012,Bias_Kacprzak_2012,Bias_Viola_2014,Bias_Kacprzak_2014}, we expect this to be less so for the LAD (Paper I).

%% LAD is a similar term as Least Squares. Does one say "the Least Squares is more sensitive..." or does one say "Least Squares [as a method] is more sensitive..."? I believe the latter is used predominantly, and "The LAD" feels odd. If the sentence should not be started by an acronym, I suggest "The method of LAD is..." or "The LAD regression..." or simply " This estimator"

\subsubsection{Efficiency} \label{sec:effi}

The formal definition of efficiency, $\tilde\eta$, relates the inherent (Fisher) information, $\mathcal{I}$, of a sample to the statistical variability around the expected value of the estimator, usually taken to be the variance, $\sigma^{2}_{\hat{\epsilon}}$, of the estimator:\begin{equation}
\label{eq:effi_fisher}
\tilde\eta = \frac{1}{\mathcal{I}\cdot\sigma^{2}_{\hat{\epsilon}}} .
\end{equation}Since the variance of an unbiased estimator cannot be less than the reciprocal of the information, $\mathcal{I}^{-1}\le\sigma^{2}_{\hat{\epsilon}}$, we have $0\le\tilde\eta\le 1$ \citep{Rao_1945,Cramer_1946}.

As we are comparing two estimators with unknown bias, it is appropriate to use the relative efficiency,\begin{equation}
\label{eq:rel_var}
\eta = \frac{\sigma^{2}_{\mu}}{\sigma^{2}_{\mathrm{LAD}}} \,,
\end{equation}with $\eta < 1$ indicating a higher efficiency for the mean, and vice versa.

Paper I showed that the LAD consistently performed better than the mean, with both higher efficiency and less bias, for various cusped intrinsic ellipticity distributions, including the shear catalog from CFHTLenS. In what follows, we take the CFHTLenS shear distribution shape to be representative of KiDS data as well since both surveys were processed with the \textsc{THELI} pipeline \citep{CFHTLenS_Erben_2013} and the shape measurement pipeline {\em lens}fit \citep{lensfit_Miller_2007,lensfit_Kitching_2008,lensfit_Miller_2013}.
% \LEt{ Consider defining.}
%Re: THELI is similar in common usage to Sextractor and is almost never defined, except perhaps in papers on the updates of the pipeline itself. See e.g. Erben et al. 2013, a reference in this paper, that discusses the usage of this pipeline, without defining the acronym.

\subsection{Halo model} \label{sec:halo_mod}

Studying the effects of estimator choice on the weak lensing signal forms the technical core of this paper. The scientific goal, however, is to assess the relevance on the inference of physical quantities, such as the derivation of a lensing halo mass from an ESD profile. Since we calculate the stacked signal for an ensemble of clusters with some common (observable) property (here a range in $r$-band luminosity), of interest is the scaling relation between the observable and derived lensing mass, $M_{200}$, where we use the definition with respect to the mean density of the universe.  

To do so, we modeled the lens density profile the same way as \citet{KiDS_Dvornik_2017}, using the halo model \citep{Halo_Seljak_2000,Halo_Peacock_2000, Halo_Cooray_2002,Halo_Bosch_2013,Halo_Cacciato_2013,Halo_Mead_2015}. The initial lens density profile is described by a Navarro-Frenk-White profile \citep{NFW}. We used the mass-concentration relation given by \citet{Halo_Duffy_2008} and allowed for a re-normalization factor, $f_{\mathrm{c}}$ \citep{KiDS_Viola_2015}.

A dominant source of systematic bias in stacked weak lensing analyses is a miscentering of the lenses, which can be due to an offset of the cluster halo with the visible distribution of galaxies \citep[see, e.g.,][]{Halo_George_2012} or the resolution of the cluster detection method \citep[less than 0.1 Mpc $h^{-1}$ for AMICO; see][]{AMICO_Bellagamba_2018}. Following \citet{Halo_Johnston_2007} as well as numerous subsequent works \citep[e.g.,][]{Halo_Oguri_2010,KiDS_Viola_2015,KiDS_Dvornik_2017,AMICO_Bellagamba_2019,Giocoli_2021_AMICO}, we allowed a fraction, $p_{\mathrm{off}}$, of clusters to be offset from the center of the galaxy distribution, effectively smoothing the central stacked $\Delta\Sigma$ profile with a characteristic radius, $\mathcal{R}_{\mathrm{off}}$.

At large radii, typically beyond a few megaparsecs, the clustering of dark matter halos starts to dominate the signal. This ``two-halo'' term depends on the halo bias, $b$ \citep{KiDS_Dvornik_2017}, and is modeled following \citet{Halo_Tinker_2010}. At small radii, the baryonic component of central galaxies can contribute to the signal, which is adequately described by a point mass, $M_{\star}$, in the model \citep{KiDS_Viola_2015,KiDS_Dvornik_2017}.
 
In Table \ref{tab:inp_priors} we summarize these six free parameters for our halo model implementation, analogous to \citet{KiDS_Dvornik_2017}.

\begin{table}[h]
\caption{Summary of the halo model fitting parameters and priors.}
\label{tab:inp_priors}
\centering
\begin{tabular}{lll}
\hline
\hline
Parameter                         &                        & Prior \\
\hline
$f_{\mathrm{c}}$                  &                        & $\left[ 0.0,8.0 \right]$ \\ 
$p_{\mathrm{off}}$                &                        & $\left[ 0.0,1.0 \right]$ \\ 
$\mathcal{R}_{\mathrm{off}}$      & [ $h^{-1}$ Mpc ]       & $\left[ 0.0,1.0 \right]$  \\ 
$b$                               &                        & $\left[ 0.0,10.0 \right]$ \\ 
$\log \left( M_{\star} \right) $  & $\log$[ $h^{-1} M_{\odot}$ ] & $\left[ 9.5,12.5 \right]$ \\ 
$\log \left( M_{200} \right) $    & $\log$[ $h^{-1} M_{\odot}$ ] & $\left[ 11.0,17.0 \right]$ \\ 
\hline                                                           
\end{tabular}                                                    
\end{table}

In the AMICO cluster sample with $6925$ lenses in $440$ square degrees, many background sources are lensed by more than one cluster, contributing to the estimate of the ESD profile in various radial bins of different clusters. In the model fitting, we took the covariance between the ESD estimates into account, as described in Sect. \ref{sec:imp_halo} \citep{KiDS_Viola_2015,KiDS_Dvornik_2017,AMICO_Bellagamba_2019}.

\section{Data and analysis}\label{sec:data}  

\begin{figure*}[h]
\centering
\resizebox{\hsize}{!}{
\includegraphics[angle=90]{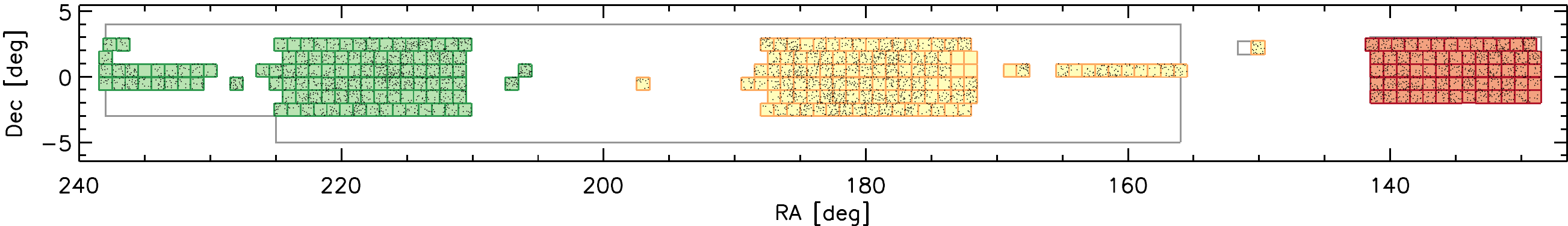}
}
\resizebox{\hsize}{!}{
\includegraphics[angle=90]{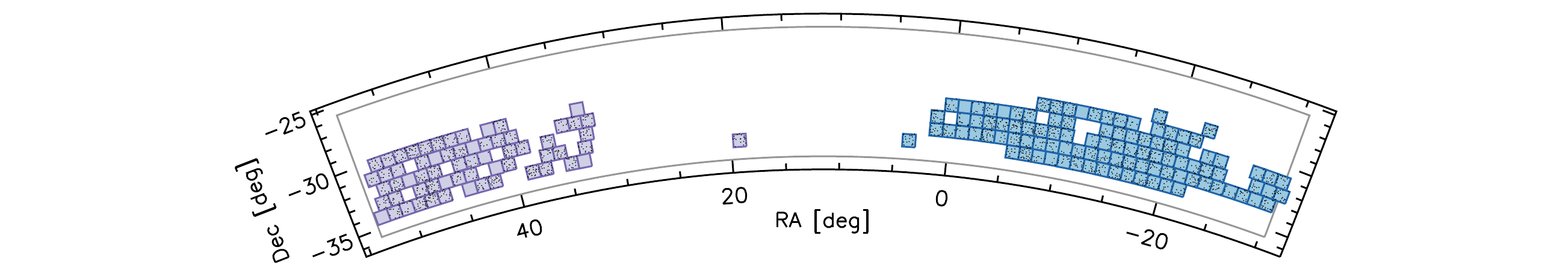}
}
\caption{Overview of the KiDS-450 observations, with the KiDS-N (upper) and KiDS-S (lower) patches. The solid gray lines represent the planned KiDS survey area. Overplotted are the observed 1 square degree tiles, color coded with respect to their correspondence with the GAMA survey patches \citep[G9 red, G12 yellow, G15 green, G23 blue, and GS purple; see][for more details]{KiDS_Hildebrandt_2017}. The AMICO clusters analyzed in this work are represented by black dots.}
\label{fig:patches}
\end{figure*}

% BB  37 -747 145 9
% BB  37 -747 183 9

% RE: \LEt{ Verify that your intended meaning has not been changed
% Note next sentence: changed "," back to "and" as these are two different catalogs
% removed "and" after KiDS-450, as KiDS itself is the suvery referred to
In this paper we use a lensing cluster catalog and a background source catalog from KiDS-450 \citep{KiDS_de_Jong_2017}. KiDS is an optical wide-field imaging survey with OmegaCAM \citep{KiDS_Kuijken_2011} on the VLT Survey Telescope \citep[VST;][]{VLT_Capaccioli_2011,KiDS_de_Jong_2013}. KiDS-450 consists of two patches, KiDS-N and KiDS-S (see Fig. \ref{fig:patches}), with 454 tiles of imaging data, for a total of 449.7 deg$^2$, in four optical filters, $ugri$. The survey was designed for lensing, ensuring a stable PSF, low seeing ($<0.96''$, with an average of $0.66''$ in $r$), and good photometric redshifts \citep[photo-$z$;][]{KiDS_Hildebrandt_2017}.

The KiDS data were reduced with \textsc{Astro-WISE} \citep{AWE_Valentijn_2007,AWE_Verdoes_2012,AWE_Begeman_2013,AWE_McFarland_2013}, as described in \citet{KiDS_de_Jong_2015,KiDS_Hildebrandt_2017}. Photometric redshifts, also termed $z_{\mathrm{B}}$, were determined using a Bayesian photo-$z$ estimation \citep[\textsc{bpz};][]{BPZ_Benitez_2000,BPZ_Coe_2006} with PSF-matched photometry, as described in \citet{CFHTLenS_Hildebrandt_2012,KiDS_Kuijken_2015,KiDS_Hildebrandt_2017}.

\subsection{Lenses} \label{sec:lenses}  

We made use of the galaxy cluster catalog derived with AMICO \citep{AMICO_Bellagamba_2011,AMICO_Radovich_2017,AMICO_Bellagamba_2018}, extracted from 440 tiles of KiDS-450 data and described in \citet{AMICO_Maturi_2019} and \citet{AMICO_Bellagamba_2019}. For each cluster, the luminosity $L_{200}$ is defined\footnote{We note that this does not take intracluster light into account.} as the sum of $r$-band luminosities of bright candidate member galaxies, weighted by membership probability \citep[see][]{AMICO_Maturi_2019}. We selected galaxies with $k$-corrected $r$-band magnitudes brighter than $m^* (z_{\mathrm{l}}) +1$ within $R_{200} (z_{\mathrm{l}})$, where $z_{\mathrm{l}}$ is the estimated cluster redshift and $R_{200}$ is derived from the adopted cluster model and is used in the construction of the cluster detection filter, as defined in \citet{AMICO_Maturi_2019}. In this sense, $L_{200}$ is defined analogously to the apparent richness, $\lambda^*$, which is a sum of membership probabilities of galaxies with $m<m^*+1.5$, within $R_{200}$.

We selected clusters in the range $0.1 \le z_{\mathrm{l}} \le 0.6$. We excluded clusters below $z=0.1$ due to their unfavorable lensing geometry and above $z=0.6$ due to the low density of background sources. For some clusters, no lens-source pairs were found, due to source selection criteria or masking. Our final selection comprises 6925 clusters, divided over the KiDS-450 survey area as shown in Fig \ref{fig:patches} and described in Table \ref{tab:KiDS-450}. The redshift distribution of these clusters is shown in Fig. \ref{fig:zdist}, with a median redshift of $z_{\mathrm{l}}=0.39$.

\begin{table}[h]
\caption{Summary of the survey patches\tablefootmark{1}, with corresponding numbers of KiDS mosaic tiles and analyzed clusters.}
\label{tab:KiDS-450}
\centering
\begin{tabular}{llll}
\hline
\hline
KiDS field & Subfield & Tiles & Clusters \\
\hline
North & G9  &  65 & 1039 \\ 
      & G12 & 113 & 1778 \\ 
      & G15 & 112 & 1737 \\ 
\hline              
South & G23 & 101 & 1517 \\ 
      & GS  &  63 &  854 \\ 
\hline                                                           
\end{tabular}
\tablefoot{
  \tablefoottext{1}{as described in \citet{KiDS_Hildebrandt_2017}}
  }
\end{table}

%; G09:         1039
%; G12:         1778 -1 masks
%; G15:         1737
%; G23:         1517 -6 -1 masks
%; GS:           854

We divided the clusters into 13 bins of cluster $L_{200}$. The limits of these bins were chosen so that the signal-to-noise ratios of the ESD measurements were approximately the same in each bin. We give an overview of these bins, together with the estimated $M_{200}$, in Table \ref{tab:clus_bins}. 
\subsection{Sources} \label{sec:sources}  

We selected an initial sample of background sources using the same photometric redshift criteria as \citet{KiDS_Hildebrandt_2017}, $0.1 < z_{\mathrm{B}} \le 0.9$, to reduce the outlier rate. We also applied  the cut $z_{\mathrm{l}}+\Delta z < z_{\mathrm{B}}$, following \citet{KiDS_Dvornik_2017}. Here, $\Delta z=0.2$ is an offset between the redshift estimation, $z_{\mathrm{l}}$, of the cluster by AMICO and the photometric redshift, $z_{\mathrm{B}}$, of the source to sufficiently lessen the contamination of the background sample by cluster member galaxies (see also Appendix \ref{app:clus_bg}).

\begin{figure}[h]
\centering
\resizebox{\hsize}{!}{
\includegraphics[angle=90]{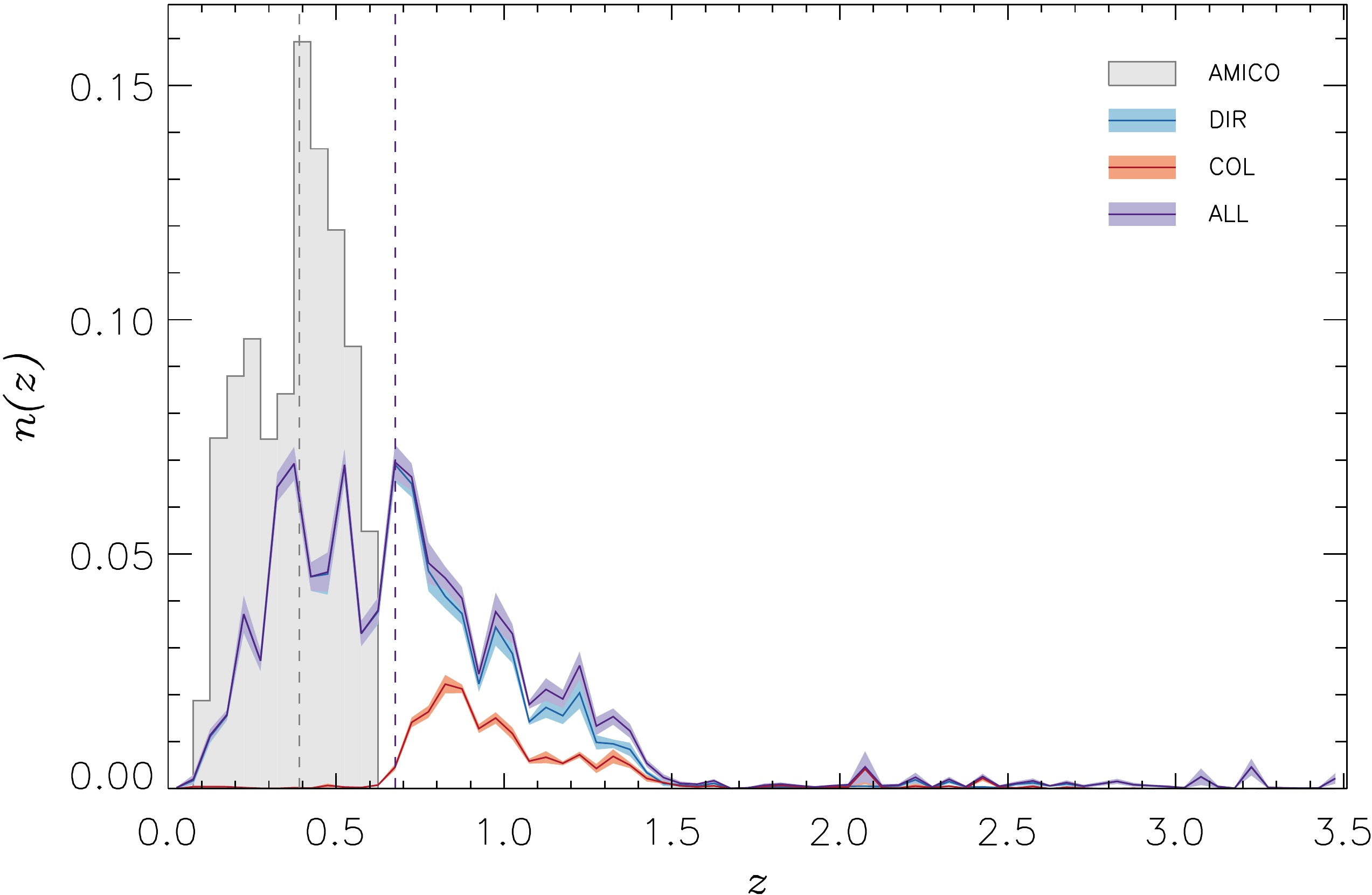}
}
\caption{Redshift distribution of AMICO clusters (gray), with a median redshift of $z_{\mathrm{l}}=0.38$ (dashed gray), and KiDS-450 background sources (purple), with a median redshift of $z=0.68$ (dashed purple). In blue (DIR), we show the initial selection following \citet{KiDS_Hildebrandt_2017} and \citet{KiDS_Dvornik_2017}. In red, we show the estimated redshift distribution of the $gri$ color selection (COL), corresponding to the bottom panel of Fig. \ref{fig:col}.}
\label{fig:zdist}
\end{figure}

\begin{figure}[h]
\centering
\resizebox{\hsize}{!}{
\includegraphics{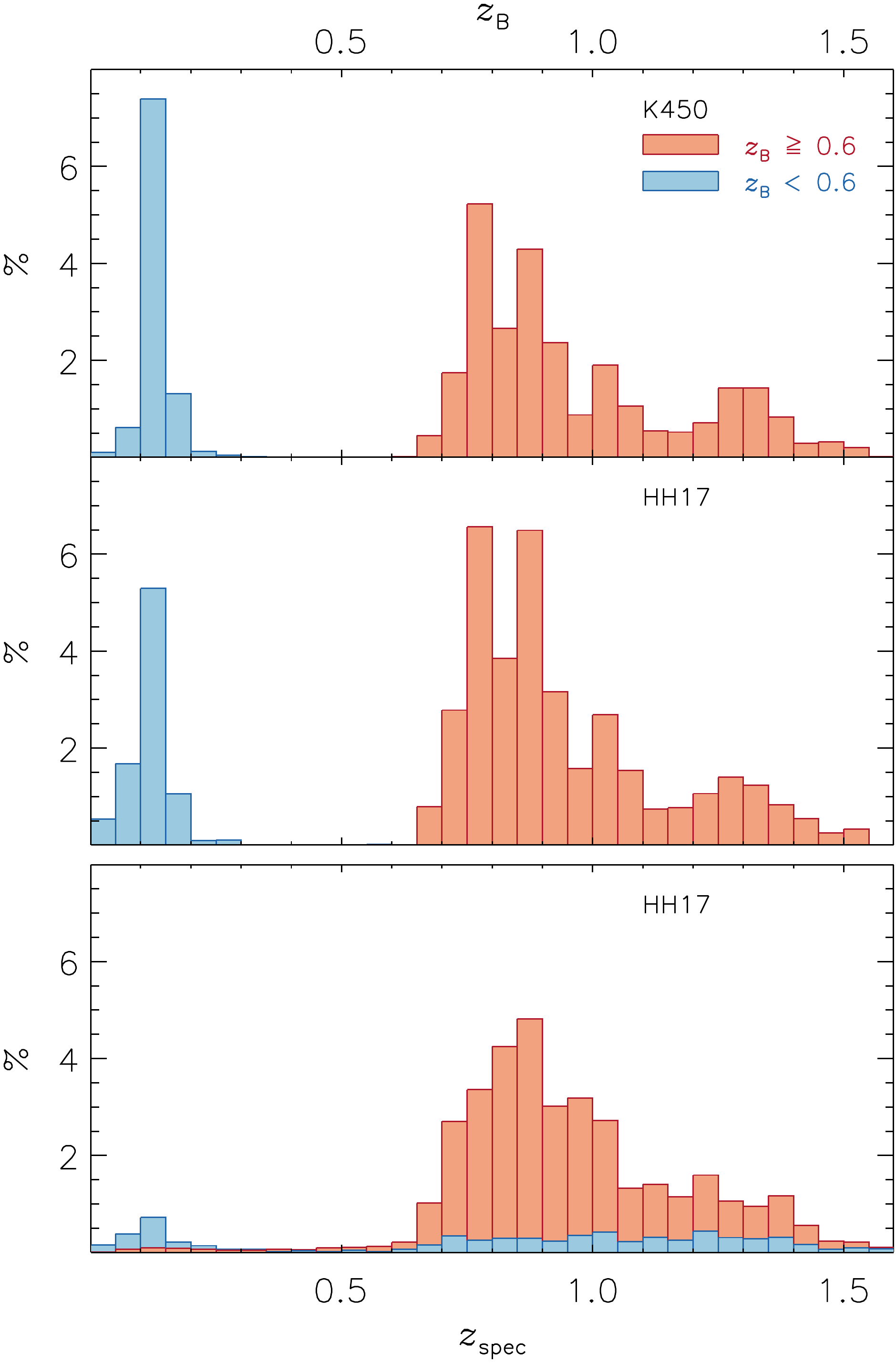}
}
\caption{Redshift distribution of background sources selected by color. The upper panel shows the distribution of the photometric redshift, $z_{\mathrm{B}}$, of the sources in the KiDS-450 catalog that satisfy the color cut of Eq.~\ref{eq:COL}, of which we select the sources with $z_{\mathrm{B}} \ge 0.6$ (red) and discard those with $z_{\mathrm{B}} < 0.6$ (blue). The bottom two panels show the same selection applied to the spec-$z$ catalog \citep{KiDS_Hildebrandt_2017}, plotted in terms of photometric redshift (middle) and spectroscopic redshift (lower). We find the contamination of sources with $z_{\mathrm{B}} \ge 0.6$ and $z_{\mathrm{spec}} < 0.6$ is $\sim 2\%$.}
\label{fig:col}
\end{figure}

Our selection of AMICO clusters is deeper than the lenses from the Galaxy And Mass Assembly \citep[GAMA][]{GAMA_Driver_2011,GAMA_Robotham_2011} catalog used in \citet{KiDS_Dvornik_2017}. As can be seen in Fig. \ref{fig:zdist}, the redshift distributions of lenses and background sources significantly overlap, and the cut at $\Delta z=0.2$ reduces the number density severely for clusters at higher redshift. Following \citet{AMICO_Bellagamba_2019}, we also selected background sources using the color selection proposed by \citet{Oguri_2012}: \begin{equation}
  \label{eq:COL}
  g-r < 0.3\quad\lor\quad r-i > 1.3 \quad\lor\quad r-i>g-r .
\end{equation}In Fig. \ref{fig:col} we show the photometric redshift distribution of this cut in the KiDS-450 catalog and compare it to the photometric and spectroscopic redshift distribution of the same cut in the spectroscopic redshift (spec-$z$) catalog used in \citet{KiDS_Hildebrandt_2017}. Based on this analysis, we additionally required $z_{\mathrm{B}} \ge 0.6$ for this selection to reduce contamination by sources at low redshift and find that 98 \% of the galaxies in this color selection have $z_{\mathrm{spec}}>0.6$.

\subsubsection{Redshift distribution} \label{sec:zdist}   

To estimate the redshift distribution of background galaxies, we did not directly use the individual redshift probability distribution, $p\left(z_{\mathrm{s}}\right),$ per source galaxy.\ Instead, we applied a weighted direct calibration method (DIR), as motivated by \citet{KiDS_Hildebrandt_2017}.

For each cluster, we used the spec-$z$ catalog described in \citet{KiDS_Hildebrandt_2017} to select objects using the same selection criteria as described above. We then used the normalized spectroscopic redshift distribution, $n\left(z_{\mathrm{s}}\right)$, of this sample to calculate the comoving critical surface density analogous to Eq. \ref{eq:sig_cr}:\begin{equation}
\label{eq:sig_cr_dir}
\left<\Sigma_{\mathrm{cr,l}}^{-1}\right> =  \frac{4\pi G}{c^2} D(z_{\mathrm{l}}) \left(1+z_{\mathrm{l}}\right)^2 \int\limits_{z_{\mathrm{l}}+\Delta z}^{\infty} \frac{D(z_{\mathrm{l}},z_{\mathrm{s}})}{D(z_{\mathrm{s}})} n(z_{\mathrm{s}})dz_{\mathrm{s}} .
\end{equation}The resulting redshift distribution for selected sources from the full KiDS-450 catalog is shown in Fig. \ref{fig:zdist}.

%BB 0 1 541 824

\subsubsection{Shape measurements} \label{sec:lensfit}  

For shape measurements, the $r$-band data were reduced using the \textsc{THELI} pipeline, developed to meet the requirements for weak gravitational lensing analyses \citep{THELI_Erben_2005,CARS_Erben_2009,THELI_Schirmer_2013,CFHTLenS_Erben_2013}. Galaxy shapes in the KiDS-450 catalog were then measured by {\em lens}fit \citep{lensfit_Miller_2007,lensfit_Kitching_2008,lensfit_Miller_2013,KiDS_Fenech_Conti_2017}.

For each source, {\em lens}fit produces the ellipticity $\left(\epsilon_1,\epsilon_2\right)$, an approximately inverse-variance weight $w_{\mathrm{s}}$ \citep[see][]{lensfit_Miller_2013}, and a fitting quality parameter. We excluded sources with unreliable ellipticities from our source sample, using the same {\em lens}fit selection criteria as described in \citet{KiDS_Hildebrandt_2017}.

In Paper I we compared the performance of estimators for ellipticity measurements in the CFHTLenS data with a subset of that catalog, selecting sources on the signal-to-noise ratio parameter $\nu_{\mathrm{SN}}$ output by {\em lens}fit. We repeated that approach for a qualitative comparison here, using two subsets of the selected KiDS-450 sources. The first set selects sources with $\nu_{\mathrm{SN}} \ge 20$, similar to Paper I, retaining $\sim 30\%$ of the full background sample. The second set is a more stringent cut of the first set, additionally selecting objects with $w_{\mathrm{s}} \ge 14.5$, comprising $\sim 20\%$ of the full sample.

\subsubsection{Effective source density} \label{sec:source_dens}  

The KiDS-450 catalog includes a filtering on general object detection and quality flags, for example, possibly blended sources or artifacts, as described by \citet{KiDS_Kuijken_2015} and \citet{KiDS_Hildebrandt_2017}, and we discarded objects that lie in a mask.\ This removed approximately $\sim 12\%$ of the sources. Our final selection comprises $14 124 197$ sources, which translates to an effective number density of $n_{\mathrm{eff}}\approx 8.23$ $\mathrm{arcmin}^{-2}$, as defined in \citet{CFHTLenS_Heymans_2012}:\begin{equation}
\label{eq:neff}
n_{\mathrm{eff}} = \frac{1}{A} \frac{\left( \sum_i w_i \right)^2}{\sum_i w_i^2} \,,
\end{equation}with $A$ the effective surface area, excluding masked regions.

% HH17 DIR: 12527622
% Our DIR: 12527616 (-6?)
% Our COL (inc overlap): 4138881
% Extra due to COL: 1596581

\subsection{Implementation} \label{sec:imp}

\subsubsection{ESD estimation}

Following \citet{AMICO_Bellagamba_2019}, we measured the ESD in 14 logarithmic bins between $0.1$ Mpc $h^{-1}$ and $3.16$ Mpc $h^{-1}$. Not only does this make for an easy comparison of the results, but it has several other practical advantages.

We avoided radii smaller than the AMICO detection pixel size, which has a median size of $0.1$  Mpc $h^{-1}$, to lessen the chance of a mismodeling the halo miscentering (Sect. \ref{sec:halo_mod}). Here, the line of sight is also most contaminated by cluster members, which can lead to an overabundance by incorrectly including ellipticity measurements that carry no lensing signal, or by an obscuring and blending of background sources, which leads to an under-abundance of sources. While these effects may partially cancel out in the number counts, the effects on the ESD measurements do not cancel out, as the first leads to a diluted signal and the second to a very poor signal-to-noise ratio (see Appendix \ref{app:clus_bg} for an assessment of cluster member contamination).

At large radii, systematic additive biases can start to play a role \citep[see, e.g.,][ for this data set]{KiDS_Dvornik_2017}, which may differ for each KiDS survey patch \citep{KiDS_Fenech_Conti_2017,KiDS_Hildebrandt_2017}. Another concern at larger separations is that the two-halo term becomes the dominant contribution to the ESD signal, which means we would need to properly constrain the halo bias, and we explain below how our approach does not fully take the clustering of dark matter halos into account.

The combination of background selection criteria from \citet{KiDS_Dvornik_2017} and \citet{AMICO_Bellagamba_2019} allows us to retain the three inner radial bins between $0.1$ and $0.2$ Mpc $h^{-1}$. We justify this inclusion in Appendix \ref{app:clus_bg}, where we repeat the tests of \citet{KiDS_Dvornik_2017}.

Each lens-source pair was then assigned a combined weight of\begin{equation}
  \label{eq:wls}
  w_{\mathrm{ls}} = w_{\mathrm{s}}\Sigma_{\mathrm{cr,l}}^{-2} \,,
\end{equation}as motivated in Sect. \ref{sec:ESD}. For LAD optimization, that is, the estimator that minimizes the $L^1$ norm (Eq. \ref{eq:loss_lad}), there exists no general analytic solution. The problem can, however, be formulated as a linear optimization, which can be solved iteratively \citep[e.g., with simplex-based methods;][]{Simplex_Barrodale_1973}. In our weak lensing analyses, we find that convergence is robust.
% \LEt{ Verify that your intended meaning has not been changed.}
% It was, reformulated

To derive the covariance matrices for the ESD estimates using the mean and LAD in the same way, we can therefore also not employ the analytical prescription of \citet{KiDS_Viola_2015} used in earlier KiDS analyses \citep[e.g.,][]{KiDS_Sifon_2015,KiDS_Uitert_2016,KiDS_Brouwer_2016}. Instead, we used a bootstrap approach.

Since the cluster bins of highest $r$-band luminosity, $L_{200}$, contain only a small number of clusters, covering only a small fraction of the KiDS-450 tiles, we cannot use the same bootstrap approach as \citet{KiDS_Viola_2015} and \citet{KiDS_Dvornik_2017} by bootstrapping 1 deg$^2$ tiles with replacement. Instead, we bootstrapped the source catalog, in accordance with \citet{AMICO_Bellagamba_2019}.

This means that we are not sensitive to the clustering effect of dark matter halos, which justifies our choice of radial lens-source separation mentioned above. To assess the accuracy of these assumptions, we estimated the covariance matrix of the full $6925$ cluster sample by bootstrapping the sources and by bootstrapping by KiDS-450 tiles in Appendix \ref{app:tile_bs}. We conclude that our bootstrapping method yields a good estimate of the covariance matrix.

\subsubsection{Halo model fitting} \label{sec:imp_halo}

Having produced the LSQ and LAD shear profiles for the stacked clusters, we fit a halo model to the results. We used the fitting procedure described in \citet{KiDS_Dvornik_2017}, producing the full posterior probabilities by a Bayesian inference technique, via a Monte Carlo Markov chain (MCMC) maximum likelihood approach. We assumed a Gaussian likelihood and made use of the full covariance between radial bins:\begin{equation}
  \mathcal{L}\propto\exp\left[-\frac{1}{2} \mathbi{R}^{\mathrm{T}}\mathbi{C}^{-1}\mathbi{R}\right] \,,
\end{equation}where the $\mathbi{R}$ are the residuals and $\mathbi{C}$ is the covariance matrix.

We used the {\tt emcee} Python package \citep{emcee_Foreman_2013} for the MCMC procedure, setting flat priors for all parameters. For the evaluation of the power spectrum and the halo mass function, we used the median redshift for each cluster luminosity bin.

\section{Results} \label{sec:results}                                             
We present our results, starting with the derived ESD profiles obtained with the mean and LAD estimators, discussing potential biases and efficiency. Then, we show the results of the halo model fitting (i.e., $M_{200}$ for each luminosity bin) and conclude with the scaling relation between $L_{200}$ and $M_{200}$. We visualize the results for the case in which all 6925 clusters are stacked together, giving the numerical results of the 13 luminosity bins in Table \ref{tab:clus_bins}. 

\subsection{ESD profiles}

We calculated the ESD profiles using $10^4$ bootstraps with replacement. We estimated the ESD signal in the 14 radial bins, using both the mean and the LAD estimators, for each bootstrapped sample, preserving the bootstrap order of all 28 values throughout the whole process.

\begin{figure}[h]
\centering
\resizebox{\hsize}{!}{
\includegraphics{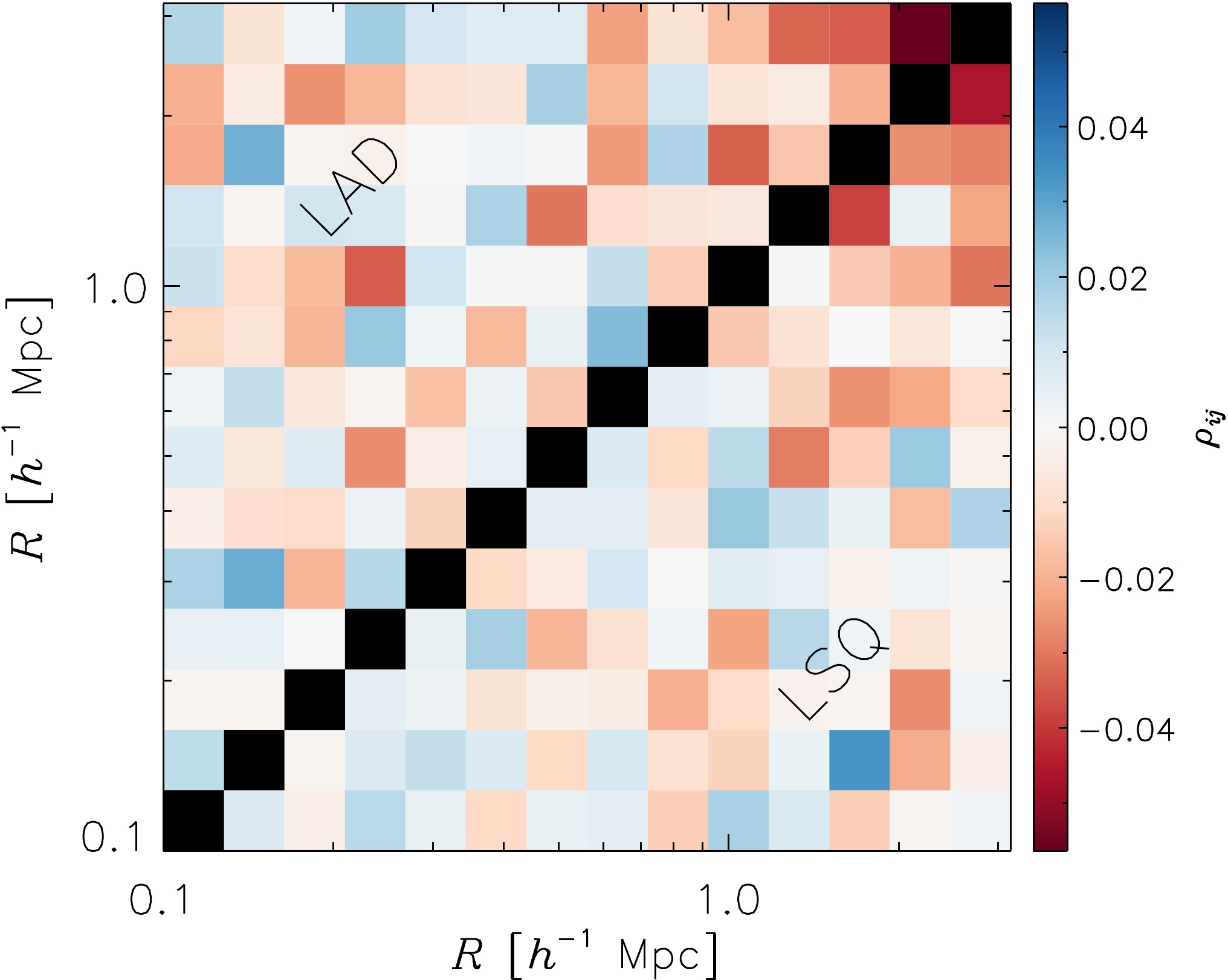}
}
\caption{Correlation matrix between the stacked ESD signals in different radial bins, using the full AMICO cluster catalog and $10^4$ bootstraps. The upper-left triangle shows the correlation, $\rho_{\mathrm{LAD}}$, between the LAD estimations. The lower-right triangle shows the correlation, $\rho_{\mathrm{mean}}$, between the ESD estimates using a weighted mean. We note the general similarities in the two patterns.}
\label{fig:COV}
\end{figure}

We find the estimator distribution to be almost perfectly normal, as expected from the central limit theorem. The correlation between the 14 bins of the full stack of clusters is shown in Fig. \ref{fig:COV} and is given by\begin{equation}
  \rho_{ij}\equiv\frac{\Cov_{ij}}{\sigma_i\sigma_j} \,,
\end{equation}where $i$ and $j$ denote the radial bin subscripts.

% BB: 4 46 564 513

The upper-left part of the matrix shows the correlation between the LAD estimates of the radial bins, and the lower-right part the shows the mean results. Although the correlation between bins is very low, it is clear that the overall trends are the same for the two estimators.

\begin{figure}[h]
\centering
\resizebox{\hsize}{!}{
\includegraphics[angle=90]{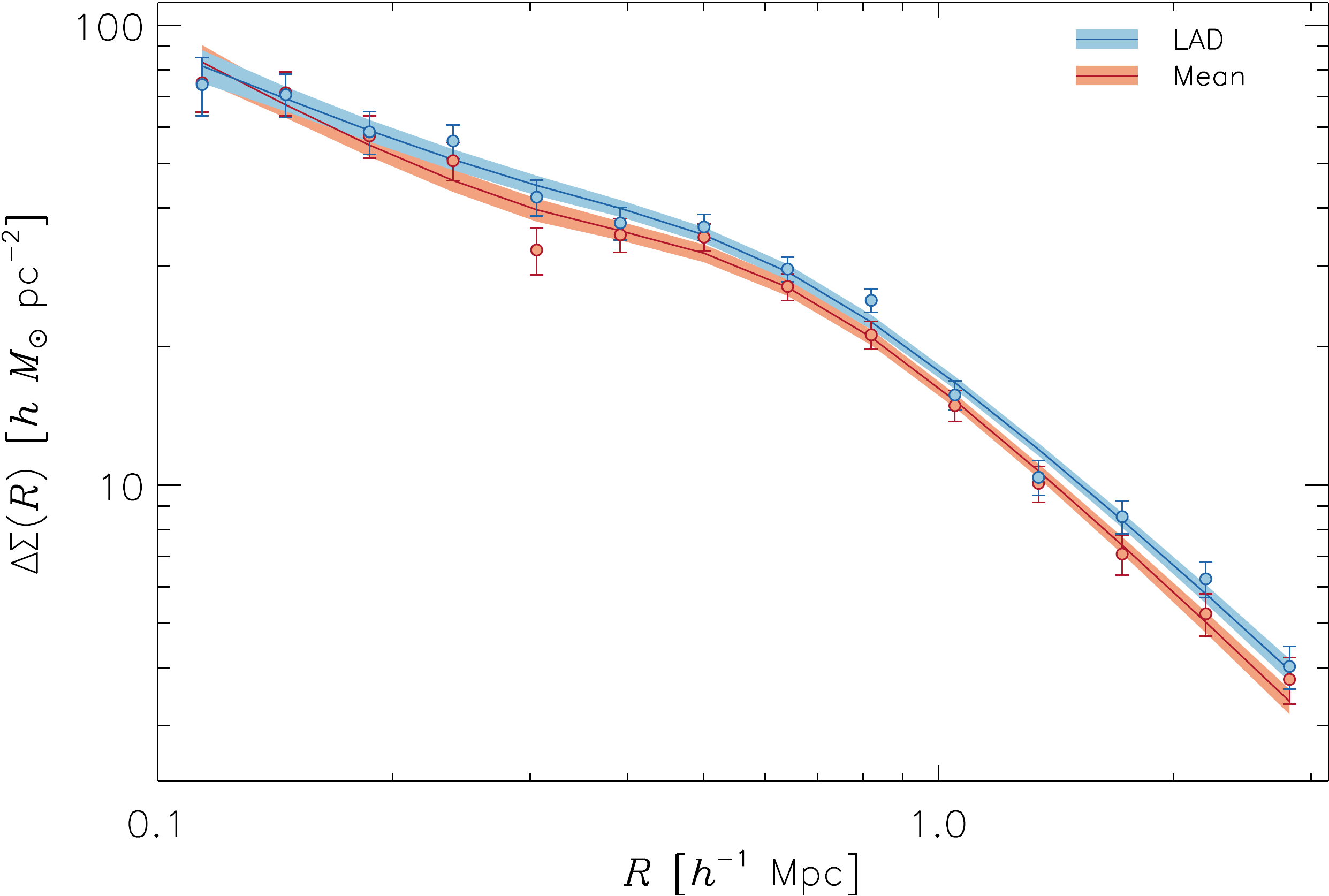}
}
\caption{Estimated ESD profile from the full AMICO cluster catalog, using the LAD estimator (blue) and a weighted mean (red). The error bars are the square roots of the diagonal values of the respective covariance matrices. The solid lines represent the best fitting halo model obtained by the MCMC fit. The shaded regions show the $68.3\%$ confidence bands, estimated using the $15.9$th and $84.1$th percentiles of the MCMC realizations.}
\label{fig:ESD}
\end{figure}

The signal-to-noise ratio of the recovered ESD profile of the full stack, which is shown in Fig. \ref{fig:ESD}, is high enough to allow us to notice the difference between the estimators, which indicates a small relative bias. The blue points show the LAD estimates, and the red points represent the mean estimates, with error bars in both cases defined as the square root of the diagonal elements of the covariance matrices (i.e., the classical standard deviation).

% BB: 46 -728 528 -14

Tests for systematic effects, such as the cross signal, and a test for systematic additive noise around random points were already conducted by \citet{KiDS_Dvornik_2017} and \citet{AMICO_Bellagamba_2019}. In Appendix \ref{app:syst} we repeat these tests for completeness since we use the KiDS-S field and an extended source selection with respect to \citet{KiDS_Dvornik_2017} and use a different source selection and three smaller radial bins with respect to \citet{AMICO_Bellagamba_2019}. Our results show no residual systematic effects, in accordance with these papers.

\subsection{Bias and efficiency} \label{res:bias_eff}

A possible bias is expected to depend on the strength of the underlying shear field since a zero lensing signal would imply no bias. In that case, the expected relevant distributions, tangential ellipticities or noise, are symmetric around zero ellipticity.

To quantify the difference between the ESD estimates, which we call the relative bias, $\Delta\Sigma_{\mathrm{Mean}} - \Delta\Sigma_{\mathrm{LAD}}$, we assumed\footnote{This assumption is only made here to quantify the bias and is not used elsewhere in the paper.} to first order\begin{equation}
  \Delta\Sigma_{\mathrm{Mean}} - \Delta\Sigma_{\mathrm{LAD}} = m \cdot \Delta\Sigma \,,
\end{equation}where we arbitrarily\footnote{We find no qualitative difference in our results when we use $\Delta\Sigma_{\mathrm{Mean}}$ instead.} use $\Sigma_{\mathrm{LAD}}$ for $\Sigma$.

\begin{figure}[h]
\centering
\resizebox{\hsize}{!}{
\includegraphics[angle=90]{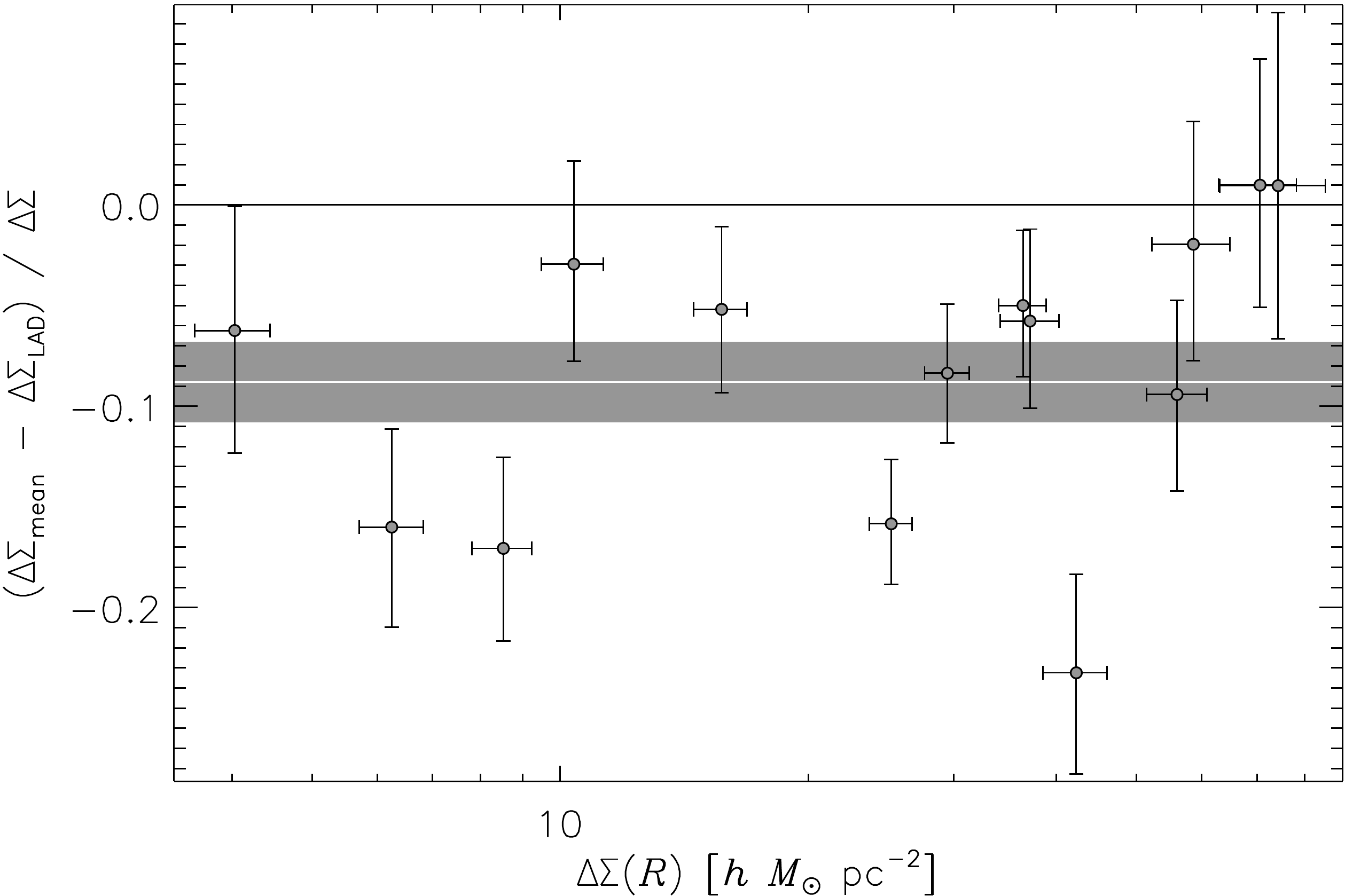}
}
\caption{Difference between the recovered ESD signals in the radial bins, showing $\left(\Delta\Sigma_{\mathrm{mean}}-\Delta\Sigma_{\mathrm{LAD}}\right) / \Delta\Sigma$. The solid white line represents the average difference, with the shaded region showing the formal 1$\sigma$ error. As a possible bias in the recovered values is expected to increase with increasing shear, the differences are plotted versus the full ESD signal in each bin where we use $\Delta\Sigma_{\mathrm{LAD}}$, but we note that the small variations in the individual points, when plotting against $\Delta\Sigma_{\mathrm{mean}}$ instead, give the same result, within statistical significance.}
\label{fig:bias_ESD}
\end{figure}

We used the full stack for its high signal-to-noise ratio, using the 15.9th and 84.1th percentiles of the differences in all bootstrap results to calculate uncertainties for each bin. We find $m = -0.088 \pm 0.020$. In Fig. \ref{fig:bias_ESD} we show this relative bias, plotting for visualization purposes\begin{equation}
  \frac{\Delta\Sigma_{\mathrm{Mean}} - \Delta\Sigma_{\mathrm{LAD}}}{\Delta\Sigma}
\end{equation}and a horizontal line at $m=-0.88$ to give a more intuitive impression of the relative error bars.

We reiterate that it is impossible to determine the absolute bias of each estimator as we did in Paper I, as we have no knowledge of the true ESD. However, the overall trend between the mean and LAD is similar in sign and order of magnitude, as we found for the CFHTLenS data in Paper I.

\begin{figure}[h]
\centering
\resizebox{\hsize}{!}{
\includegraphics[angle=90]{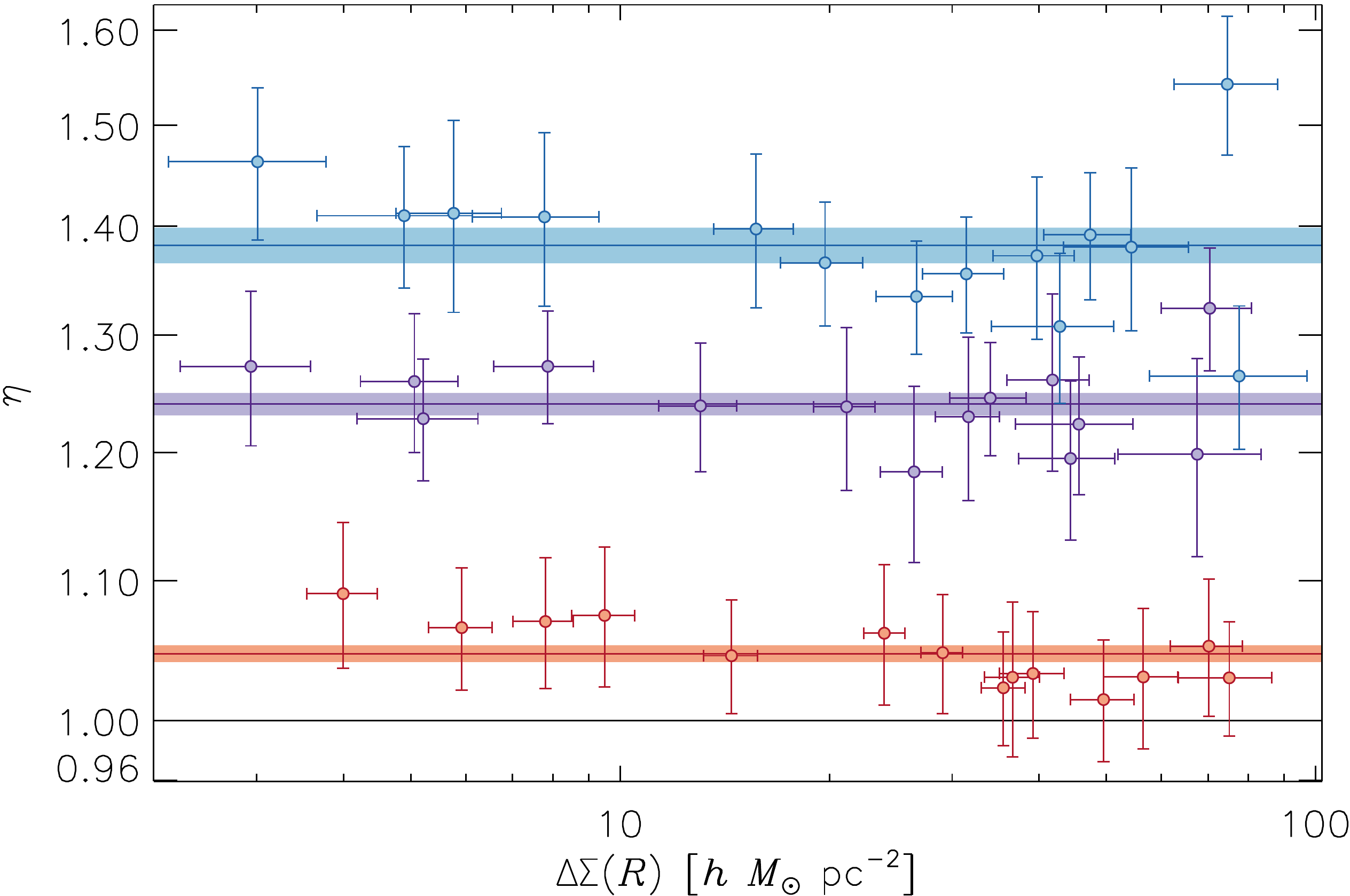}
}
\caption{Relative efficiencies, $\eta\equiv\sigma^2_{\mathrm{Mean}}/\sigma^2_{\mathrm{LAD}}$, defined as the ratio of the diagonal elements of the covariance matrices. The relative efficiencies using the full KiDS-450 source catalog are shown in red. The purple and blue represent the higher S/N and higher {\em lens}fit weight selections, respectively. The solid lines represent the average $\eta$ for each selection, with the shaded regions showing the 1$\sigma$ errors.}
\label{fig:eff_ESD}
\end{figure}

In Fig. \ref{fig:eff_ESD} we show the derived relative efficiency (Eq. \ref{eq:rel_var}) $\eta = 1.047 \pm 0.006$, which is in accordance with the findings for CFHTLenS data in Paper I. The measured\footnote{i.e., the combination of the intrinsic distribution and the various effects before and after the lensing by AMICO clusters, which affects the observation and measurement of the source ellipticities.} ellipticity distribution is expected to differ for shape measurements with a higher signal-to-noise ratio. For example, the cuspiness of the distribution shown in Fig.~\ref{dist_kids} can be smoothed out by noise convolutions. As the LAD estimation is more sensitive to the central peak, this will affect its precision.

In Paper I this was confirmed in the comparison of simulated data with and without noise, as well as in the results of the CFHTLenS sample with a stringent signal-to-noise selection compared with the full sample. As in Paper I, we compared the relative efficiencies for our selections with $\nu_{\mathrm{SN}}\ge 20$ and with $w_i\ge 14.5$, finding indeed a higher efficiency for less noisy shapes, namely $\eta = 1.240 \pm 0.010$ and $\eta = 1.386 \pm 0.018$, respectively.

\subsection{Halo masses} \label{sec:halo_res}                                             
                                                                 
We ran MCMC chains of 120 000 samples, using 120 walkers with 1000 steps each. The resulting chains were fully converged after the first 200 steps, so we discarded the first 24 000 samples.

We summarize the $M_{200}$ derived from the ESD estimation of the 13 luminosity bins in Table \ref{tab:clus_bins}. Reduced $\chi^2$, estimated between 0.730 and 2.528, are fairly consistent between derived results for mean and LAD.

\begin{table*}[h]
\caption{Properties and lensing results of the individual luminosity bins.}
\label{tab:clus_bins}
\centering
\begin{tabular}{ccrccccc}
\hline
\hline
Luminosity bin  & Median $L_{200}$                       & Clusters & Median $z_{\mathrm{l}}$ & $M_{200}$ Mean                          & $\chi^2_\nu$ & $M_{200}$ LAD                          & $\chi^2_\nu$ \\
$\left[ 10^{10}h^{-2}L_{\odot} \right]$ & $\left[ 10^{10}h^{-2}L_{\odot} \right]$ &          &                        & $\left[ 10^{14}h^{-1}M_{\odot} \right]$  &              & $\left[ 10^{14}h^{-1}M_{\odot} \right]$ &             \\
\hline
 $[0.4,17.4[$   &  $12.3^{+3.5}_{-4.6}$ &  2346 & 0.29  & $0.187^{+0.029}_{-0.026}$ & 1.27 & $0.191^{+0.031}_{-0.026}$ & 1.18 \\ 
 $[17.4,24.8[$  &  $20.8^{+2.7}_{-2.3}$ &  1545 & 0.41  & $0.416^{+0.061}_{-0.076}$ & 2.19 & $0.445^{+0.055}_{-0.061}$ & 1.60 \\ 
 $[24.8,31.8[$  &  $28.0^{+2.5}_{-2.2}$ &  1027 & 0.41  & $0.371^{+0.048}_{-0.046}$ & 0.81 & $0.385^{+0.050}_{-0.046}$ & 0.91 \\ 
 $[31.8,40.5[$  &  $35.2^{+3.2}_{-2.4}$ &  685  & 0.42  & $0.519^{+0.078}_{-0.078}$ & 1.34 & $0.621^{+0.076}_{-0.074}$ & 1.29 \\ 
 $[40.5,49.0[$  &  $44.4^{+2.9}_{-2.7}$ &  457  & 0.42  & $1.076^{+0.182}_{-0.181}$ & 2.18 & $0.998^{+0.150}_{-0.139}$ & 2.53 \\ 
 $[49.0,59.9[$  &  $54.1^{+3.6}_{-3.9}$ &  305  & 0.40  & $1.387^{+0.425}_{-0.335}$ & 1.05 & $1.462^{+0.533}_{-0.353}$ & 0.98 \\ 
 $[59.9,72.9[$  &  $65.2^{+5.2}_{-3.7}$ &  202  & 0.41  & $1.318^{+0.267}_{-0.234}$ & 0.77 & $1.309^{+0.233}_{-0.203}$ & 0.74 \\ 
 $[72.9,84.1[$  &  $78.6^{+3.2}_{-3.9}$ &  135  & 0.38  & $1.406^{+0.344}_{-0.215}$ & 1.22 & $1.528^{+0.409}_{-0.254}$ & 1.30 \\ 
 $[84.1,102[$   &  $91.8^{+5.7}_{-4.8}$ &  90   & 0.39  & $2.438^{+0.486}_{-0.443}$ & 0.75 & $2.472^{+0.479}_{-0.425}$ & 0.73 \\ 
 $[102,129[$    &  $112^{+10}_{-7.2}$ &  60   & 0.40  & $2.143^{+0.618}_{-0.417}$ & 0.77 & $1.914^{+0.680}_{-0.373}$ & 0.79 \\ 
 $[129,160[$    &  $138^{+11}_{-6.9}$ &  40   & 0.395 & $3.999^{+1.957}_{-0.993}$ & 1.35 & $3.715^{+1.557}_{-0.910}$ & 1.64 \\ 
 $[160,221[$    &  $175^{+27}_{-11}$ &  26   & 0.37  & $4.207^{+0.713}_{-0.610}$ & 1.47 & $4.786^{+0.824}_{-0.637}$ & 1.18 \\ 
 $[221,400[$    &  $277^{+106}_{-24}$ &  8    & 0.375 & $7.638^{+2.293}_{-1.613}$ & 1.02 & $9.141^{+3.105}_{-1.913}$ & 0.81 \\ 
\hline                                                           
\end{tabular}
\tablefoot{
  The errors on the median $L_{200}$ and derived $M_{200}$ in each luminosity bin are the differences with the 15.9th and 84.1th percentiles.
  }
\end{table*}

For the full stack of clusters, we derived\begin{subequations}
  \label{eq:M200_stack}
  \begin{align}
    M_{200} = \left(0.453^{+0.030}_{-0.030}\right) \times 10^{14} h^{-1} M_{\odot} \,, & \quad \chi^2_\nu=1.25 \; & \mathrm{(Mean)} \\
   M_{200} =  \left(0.487^{+0.033}_{-0.036}\right) \times 10^{14} h^{-1} M_{\odot} \,, & \quad \chi^2_\nu=1.37 \; & \mathrm{(LAD).}
  \end{align}
\end{subequations}The confidence intervals are derived from the 15.9th and 84.1th percentiles of the posterior distributions. The best fitting ESD models are shown in Fig. \ref{fig:ESD}. The $68.3\%$ confidence bands overlap at some radii and are in tension at other radii. While the difference in ESD is significant, the $68.3\%$ confidence intervals for $M_{200}$ just touch.

%      1.24633      1.36917
%    -0.303735      4.52876    -0.303199
%    -0.356333      4.87077    -0.327976

\subsection{$L_{200}-M_{200}$ scaling relation} \label{res:scaling}

We assumed a power-law relation between the derived halo masses and the median $r$-band luminosity of each cluster bin. We fit this relation in the form\begin{equation}
\log\left(\frac{M_{200}}{M_{\mathrm{piv}}}\right)=a+b\log\left(\frac{L_{200}}{L_{\mathrm{piv}}}\right) \,,
\end{equation}with $a$ the intercept and $b$ the slope, where $M_{\mathrm{piv}} \approx 10^{14.1} h^{-1} M_{\odot}$ and $L_{\mathrm{piv}} \approx 10^{11.8} h^{-2} L_{\odot}$ are typical pivotal values of the halo mass and luminosity, derived from the fit itself. The fit was done in log basis as the derived posterior distributions of the halo mass are log-normal.

%We do not only minimize the residuals between the halo masses and the scaling relation, but also explicitly take the spread of $r$-band luminosities into account, as we are fitting a relation between two quantities with a non-negligible variability.

We did not take a redshift dependence into account, as \citet{AMICO_Bellagamba_2019} showed only a marginal and not very steep dependence of the halo mass on redshift.

\begin{figure}[h]
\centering
\resizebox{\hsize}{!}{
\includegraphics[angle=90]{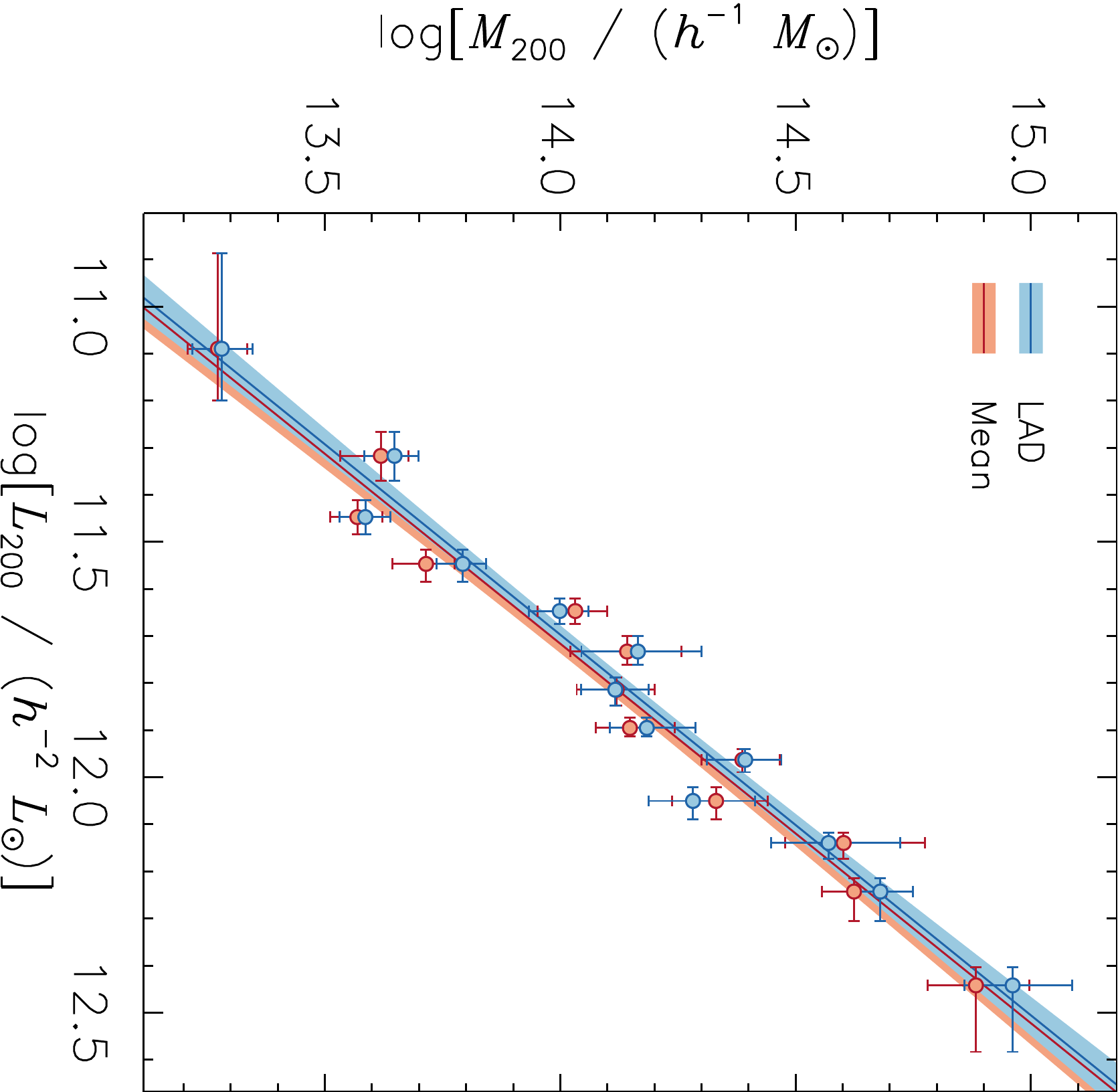}
}
\caption{$r$-band luminosity-halo mass scaling relations, derived from the ESD profiles estimated using the weighted mean (red) and LAD (blue). At the pivot point, recognizable as the narrowest parts of the confidence bands, the relations just touch at the $68.3\%$ confidence level.}
\label{fig:powerlaw}
\end{figure}

We obtained the scaling relations\begin{subequations}
  \label{eq:powerlaw}
  \begin{align}
     \frac{M_{200}}{10^{14.1} h^{-1} M_{\odot}} = (0.97 \pm 0.06) \left( \frac{L_{200}}{10^{11.8} h^{-2} L_{\odot}} \right)^{(1.24 \pm 0.08)} & \mathrm{(Mean)} \\
     \frac{M_{200}}{10^{14.1} h^{-1} M_{\odot}} = (1.03 \pm 0.05) \left( \frac{L_{200}}{10^{11.8} h^{-2} L_{\odot}} \right)^{(1.24 \pm 0.08)} & \mathrm{(LAD)}
  \end{align}
\end{subequations}and plot the results in Fig. \ref{fig:powerlaw}. As with the derived ESD profiles and halo masses for the full stack of clusters, the $68.3\%$ confidence bands just touch at the pivot point\begin{equation}
  \left(L_{200},M_{200}\right) = \left(10^{11.8} h^{-2} L_{\odot},10^{14.1} h^{-1} M_{\odot}\right) \,,
\end{equation}recognizable as the narrowest parts of the confidence bands.

% BB: 46 -476 528 -6

% BB 46 -476 528 -6

% Confidence
%
% 0.92  -0.06   0.98   0.06   1.04
% 0.98  -0.05   1.03   0.05   1.09

\section{Summary and conclusions} \label{Summary}

% preamble
% science results

We conducted a weak shear analysis of 6925 AMICO clusters in the KiDS-450 data We derived a tightly constrained scaling relation between $r$-band luminosity, $L_{200}$, and average lensing masses, $M_{200}$, in concordance with earlier results in the literature.
% preamble
% estimator performance

We investigated the impact of estimator choice for inferring the central moment of the cusped and skewed ellipticity distribution of background galaxies, finding a relative bias on the order of a few percent, as predicted in Paper I. We find that the constraints obtained via LAD regression are tighter than those obtained via LSQ regression, and they significantly improved as the signal-to-noise ratio of the shape measurements of the background galaxies increased. Complemented by simulations from Paper I, we give an alternative perspective on the problem of inferring the central shear value from the skewed distribution of background galaxy shapes, at the minor cost of increased, but still feasible, computation times for numerical iterative regression.
% \LEt{ Verify that your intended meaning has not been changed.}
% It's fine

\subsection{$L_{200}-M_{200}$ relation}

% main conclusion: bias approx linear with ESD, so expected similar slope and different bias

Since the relative bias we found in both this research and in Paper I is approximately proportional to the ESD signal, it is expected that the LAD estimator will mainly have an effect on the intercept of the $L_{200}-M_{200}$ scaling relation. This was confirmed by our results in Eq. \ref{eq:powerlaw}.
%\LEt{ single-sentence paragraph.}
% Reformulated

% power-law index and consistency with literature

The power-law index of the $L_{200}-M_{200}$ scaling relation was constrained to $1.24\pm 0.08$ (Eq. \ref{eq:powerlaw}), independent of estimator choice. This is in agreement with earlier work in the literature, such as \citet[][and references therein]{KiDS_Viola_2015}, who cite $1.16\pm 0.13$. This agreement is noteworthy since the AMICO clusters are derived from photometric redshifts, as opposed to the spectroscopically derived groups from GAMA \citep{GAMA_Driver_2011,GAMA_Robotham_2011}.  The difference in confidence is explained by the increased number of lenses, which is not surprising given the large overlap in mass range: \citet{KiDS_Viola_2015} analyzed 1413 galaxy groups between $z=0.03$ and $z=0.33$, with $r$-band luminosity bin limits between $2.5\times 10^9$ and $5.0\times 10^{12}$ $h^{-2} L_{\odot}$, deriving halo masses between $1.4\times 10^{13}$ and $3.5\times 10^{14}$ $h^{-1} M_{\odot}$.

% intercept and relative bias between estimators

The choice of estimator produces a difference in intercept at just the $1\sigma$ level. Using a weighted mean to derive the ESD leads to an intercept of $0.97 \pm 0.06$ in the scaling relation, while LAD gives an intercept of $1.03 \pm 0.05$. This is to be expected as the relative bias seems roughly constant, when normalized by the ESD (see Fig. \ref{fig:bias_ESD}), and in good agreement with the bias of $\sim 5\%$ found using simulations in Paper I.

% Further evaluation of results

As in \citet{AMICO_Bellagamba_2019}, we estimate that systematic effects mostly affect the intercept. While the derived intercept is in agreement with the aforementioned papers, we note that the chosen definition\footnote{e.g., \citet{KiDS_Viola_2015}, where the group $r$-band luminosities are calculated by summing over spectroscopically confirmed group members.} of $L_{200}$, the difference in redshift range and definitions, and the completeness of group and cluster membershipcan account for possible differences on the same order of magnitude. This would not affect our conclusions on methodology, as the results from both estimators would be similarly affected.
% \LEt{ Verify that your intended meaning has not been changed.}

In a further comparison with the scaling relation between richness and mass, cited in \citet{AMICO_Bellagamba_2019}, we find similar significance in constraints on the slope. We define bins in luminosity instead of AMICO detection amplitude \citep{AMICO_Radovich_2017} or richness, but since the cluster luminosity is tightly correlated with the richness, $\lambda^*$, this confirms our findings.

There are a few differences to consider. We chose not to account for a possible redshift dependence. This is motivated by \citet{AMICO_Bellagamba_2019} finding only a shallow dependence, which they point out may be driven mainly by the highest redshift bin. Our lensing analysis employs a slightly different background selection for an increased source density, combined with a different derivation of the associated redshift distribution, $n(z)$. Another difference is the inclusion of radial bins at $0.1 \le R < 0.2$ Mpc $h^{-1}$. This is expected to only have a minor effect as the contribution of the stellar mass is an order of magnitude lower than the halo term at these radii, while the contribution of miscentered halos only starts to become significant at larger radii \citep[see also][]{Halo_Rykoff_2016,Halo_Oguri_2018}. In this sense, these findings are a confirmation of the robustness of the results across these papers.

\subsection{Optimal estimators}

Our results are in good agreement with Paper I, with a relative bias between the two estimators that shows the recovered lensing signal is higher with LAD, suggestive of a lower absolute bias. At the same time, LAD regression gives a small (albeit significant) gain in efficiency, giving a reduction in error bars of a few percent, and potentially up to $11\% - 18\%$ for shape measurements of a higher signal-to-noise ratio. Least absolute deviation regression comes at the cost of a higher computation time, but at a step in the analysis process that does not dominate the total computational cost.

Both simulations (Paper I) and analyses on real data (this article) cite quantitative results of significance while at the same time showing similar trends between estimators on a qualitative level. We have conducted a cautious and thorough investigation but can never exclude the unknown: biases arising due to assumptions in the simulations of Paper I or uncorrected systematic effects in this research, or, most likely, both. However, given the range and realism in simulated distributions and the similarities in findings among those simulations, this research, and other work in the literature, we are confident that the recovered differences in results between the two estimators are real. 

We note some differences between the two analyses. In Paper I we analyzed the regression of a sample of ellipticities with a single underlying value of the shear and, for each type of simulation, a single intrinsic ellipticity distribution. In this research, the situation is more complex. We studied the stacked signal around samples of lenses and of samples of background sources at a range of redshifts. This means also stacking noise that has been scaled by a range in lensing geometries, quantified by $\Sigma^{-2}_{\mathrm{cr}}$. Furthermore, in each radial bin of each luminosity bin, we assume: (i) a constant lensing effect, which is in reality the stacked average of a range in $L_{200}$, and therefore a range in $M_{200}$, confounded by intrinsic scatter between these two quantities, and (ii) radial distance $R$ from the lens, combined with a miscentering of halos.
%\LEt{ Verify that your intended meaning has not been changed.}.

All these effects tend to convolve the intrinsic galaxy shape distribution, which makes the level of agreement and significance between the two papers in fact remarkably robust. In conclusion:\begin{itemize}
    \item The combination of Paper I and this research shows that LAD regression is more naturally suited to the cusped intrinsic ellipticity distribution of background galaxies.
    \item Our simulations in Paper I showed a lower bias for LAD regression than for LSQ regression in the presence of noise in the background source shape measurements, while this research confirmed the same relative bias between the two estimators.
    \item Constraints obtained via LAD regression are comparable with or tighter than constraints obtained via LSQ regression.
\end{itemize}An optimal estimator is, from a principled point of view, more objective and better suited than corrections to an approach, which is known to mismatch the sample distribution. More practically, LAD regression provides a robust consistency check for shear inference, which has been and still remains a major investment in the field of weak lensing. %Reiteration our starting point: whether comparing data to simulations or calibrating data with deeper observations of higher resolution, one must be as certain as possible of what one is looking at. 
Keeping different perspectives, such as exploring these alternative statistical approaches, is fundamental for determining the way forward.

%\subsection{Future work}

\begin{acknowledgements}

MSm acknowledges support from the Netherlands Organization for Scientific Research (NWO).

AD acknowledges ERC Consolidator Grant (No. 770935)

LM acknowledges the grants ASI-INAF n. 2018-23-HH.0 and PRIN-MIUR 2017 WSCC32 “Zooming into dark matter and proto-galaxies with massive lensing clusters”.

MSe acknowledges financial contribution from contract ASI-INAF n.2017-14-H.0 and contract INAF mainstream project 1.05.01.86.10. 

Based on data products from observations made with ESO Telescopes at the La Silla Paranal Observatory under programme IDs 177.A-3016, 177.A-3017 and 177.A-3018, and on data products produced by Target/OmegaCEN, INAF-OACN, INAF-OAPD and the KiDS production team, on behalf of the KiDS consortium. OmegaCEN and the KiDS production team acknowledge support by NOVA and NWO-M grants. Members of INAF-OAPD and INAF-OACN also acknowledge the support from the Department of Physics \& Astronomy of the University of Padova, and of the Department of Physics of Univ. Federico II (Naples).

\end{acknowledgements}

\bibliographystyle{aa}
\bibliography{Clusters_KiDS_stats.bib}

\begin{thebibliography}{115}
\expandafter\ifx\csname natexlab\endcsname\relax\def\natexlab#1{#1}\fi

\bibitem[{{Bacon} {et~al.}(2000){Bacon}, {Refregier}, \&
  {Ellis}}]{WLhist_Bacon_2000}
{Bacon}, D.~J., {Refregier}, A.~R., \& {Ellis}, R.~S. 2000, \mnras, 318, 625

\bibitem[{{Barrodale} \& {Roberts}(1973)}]{Simplex_Barrodale_1973}
{Barrodale}, I. \& {Roberts}, F.~D.~K. 1973, SIAM Journal on Numerical
  Analysis, 10, 839

\bibitem[{{Bartelmann} \& {Maturi}(2017)}]{WLgen_BM_2017}
{Bartelmann}, M. \& {Maturi}, M. 2017, Scholarpedia, 12, 32440

\bibitem[{{Bartelmann} \& {Schneider}(2001)}]{WLgen_BS_2001}
{Bartelmann}, M. \& {Schneider}, P. 2001, \physrep, 340, 291

\bibitem[{{Begeman} {et~al.}(2013){Begeman}, {Belikov}, {Boxhoorn}, \&
  {Valentijn}}]{AWE_Begeman_2013}
{Begeman}, K., {Belikov}, A.~N., {Boxhoorn}, D.~R., \& {Valentijn}, E.~A. 2013,
  Experimental Astronomy, 35, 1

\bibitem[{{Bellagamba} {et~al.}(2011){Bellagamba}, {Maturi}, {Hamana},
  {Meneghetti}, {Miyazaki}, \& {Moscardini}}]{AMICO_Bellagamba_2011}
{Bellagamba}, F., {Maturi}, M., {Hamana}, T., {et~al.} 2011, \mnras, 413, 1145

\bibitem[{{Bellagamba} {et~al.}(2018){Bellagamba}, {Roncarelli}, {Maturi}, \&
  {Moscardini}}]{AMICO_Bellagamba_2018}
{Bellagamba}, F., {Roncarelli}, M., {Maturi}, M., \& {Moscardini}, L. 2018,
  \mnras, 473, 5221

\bibitem[{{Bellagamba} {et~al.}(2019){Bellagamba}, {Sereno}, {Roncarelli},
  {Maturi}, {Radovich}, {Bardelli}, {Puddu}, {Moscardini}, {Getman},
  {Hildebrandt}, \& {Napolitano}}]{AMICO_Bellagamba_2019}
{Bellagamba}, F., {Sereno}, M., {Roncarelli}, M., {et~al.} 2019, \mnras, 484,
  1598

\bibitem[{{Ben{\'\i}tez}(2000)}]{BPZ_Benitez_2000}
{Ben{\'\i}tez}, N. 2000, \apj, 536, 571

\bibitem[{{Bernstein}(2010)}]{SYST_Bernstein_2010}
{Bernstein}, G.~M. 2010, \mnras, 406, 2793

\bibitem[{{Bernstein} \& {Armstrong}(2014)}]{Shape_Bernstein_2014}
{Bernstein}, G.~M. \& {Armstrong}, R. 2014, \mnras, 438, 1880

\bibitem[{{Bernstein} \& {Jarvis}(2002)}]{Shape_Bernstein_2002}
{Bernstein}, G.~M. \& {Jarvis}, M. 2002, \aj, 123, 583

\bibitem[{{Bonnet} \& {Mellier}(1995)}]{WLhist_Bonnet_1995}
{Bonnet}, H. \& {Mellier}, Y. 1995, \aap, 303, 331

\bibitem[{{Brainerd} {et~al.}(1996){Brainerd}, {Blandford}, \&
  {Smail}}]{WLhist_Brainerd_1996}
{Brainerd}, T.~G., {Blandford}, R.~D., \& {Smail}, I. 1996, \apj, 466, 623

\bibitem[{{Bridle} {et~al.}(2010){Bridle}, {Balan}, {Bethge}, {Gentile},
  {Harmeling}, {Heymans}, {Hirsch}, {Hosseini}, {Jarvis}, {Kirk}, {Kitching},
  {Kuijken}, {Lewis}, {Paulin-Henriksson}, {Sch{\"o}lkopf}, {Velander},
  {Voigt}, {Witherick}, {Amara}, {Bernstein}, {Courbin}, {Gill}, {Heavens},
  {Mandelbaum}, {Massey}, {Moghaddam}, {Rassat}, {R{\'e}fr{\'e}gier}, {Rhodes},
  {Schrabback}, {Shawe-Taylor}, {Shmakova}, {van Waerbeke}, \&
  {Wittman}}]{GREAT_Bridle_2010}
{Bridle}, S., {Balan}, S.~T., {Bethge}, M., {et~al.} 2010, \mnras, 405, 2044

\bibitem[{{Brouwer} {et~al.}(2016){Brouwer}, {Cacciato}, {Dvornik}, {Eardley},
  {Heymans}, {Hoekstra}, {Kuijken}, {McNaught-Roberts}, {Sif{\'o}n}, {Viola},
  {Alpaslan}, {Bilicki}, {Bland-Hawthorn}, {Brough}, {Choi}, {Driver}, {Erben},
  {Grado}, {Hildebrandt}, {Holwerda}, {Hopkins}, {de Jong}, {Liske},
  {McFarland}, {Nakajima}, {Napolitano}, {Norberg}, {Peacock}, {Radovich},
  {Robotham}, {Schneider}, {Sikkema}, {van Uitert}, {Verdoes Kleijn}, \&
  {Valentijn}}]{KiDS_Brouwer_2016}
{Brouwer}, M.~M., {Cacciato}, M., {Dvornik}, A., {et~al.} 2016, \mnras, 462,
  4451

\bibitem[{{Cacciato} {et~al.}(2013){Cacciato}, {van den Bosch}, {More}, {Mo},
  \& {Yang}}]{Halo_Cacciato_2013}
{Cacciato}, M., {van den Bosch}, F.~C., {More}, S., {Mo}, H., \& {Yang}, X.
  2013, \mnras, 430, 767

\bibitem[{{Capaccioli} \& {Schipani}(2011)}]{VLT_Capaccioli_2011}
{Capaccioli}, M. \& {Schipani}, P. 2011, The Messenger, 146, 2

\bibitem[{{Clampitt} \& {Jain}(2016)}]{Halo_Clampitt_2016}
{Clampitt}, J. \& {Jain}, B. 2016, \mnras, 457, 4135

\bibitem[{{Coe} {et~al.}(2006){Coe}, {Ben{\'\i}tez}, {S{\'a}nchez}, {Jee},
  {Bouwens}, \& {Ford}}]{BPZ_Coe_2006}
{Coe}, D., {Ben{\'\i}tez}, N., {S{\'a}nchez}, S.~F., {et~al.} 2006, \aj, 132,
  926

\bibitem[{{Cooray} \& {Sheth}(2002)}]{Halo_Cooray_2002}
{Cooray}, A. \& {Sheth}, R. 2002, \physrep, 372, 1

\bibitem[{Cramer(1946)}]{Cramer_1946}
Cramer, H. 1946, Mathematical Methods of Statistics (Princeton mathematical
  series, Princeton University Press)

\bibitem[{{Dark Energy Survey Collaboration} {et~al.}(2016){Dark Energy Survey
  Collaboration}, {Abbott}, {Abdalla}, {Aleksi{\'c}}, {Allam}, {Amara},
  {Bacon}, {Balbinot}, {Banerji}, {Bechtol}, {Benoit-L{\'e}vy}, {Bernstein},
  {Bertin}, {Blazek}, {Bonnett}, {Bridle}, {Brooks}, {Brunner}, {Buckley-Geer},
  {Burke}, {Caminha}, {Capozzi}, {Carlsen}, {Carnero-Rosell}, {Carollo},
  {Carrasco-Kind}, {Carretero}, {Castander}, {Clerkin}, {Collett}, {Conselice},
  {Crocce}, {Cunha}, {D'Andrea}, {da Costa}, {Davis}, {Desai}, {Diehl},
  {Dietrich}, {Dodelson}, {Doel}, {Drlica-Wagner}, {Estrada}, {Etherington},
  {Evrard}, {Fabbri}, {Finley}, {Flaugher}, {Foley}, {Fosalba}, {Frieman},
  {Garc{\'\i}a-Bellido}, {Gaztanaga}, {Gerdes}, {Giannantonio}, {Goldstein},
  {Gruen}, {Gruendl}, {Guarnieri}, {Gutierrez}, {Hartley}, {Honscheid}, {Jain},
  {James}, {Jeltema}, {Jouvel}, {Kessler}, {King}, {Kirk}, {Kron}, {Kuehn},
  {Kuropatkin}, {Lahav}, {Li}, {Lima}, {Lin}, {Maia}, {Makler}, {Manera},
  {Maraston}, {Marshall}, {Martini}, {McMahon}, {Melchior}, {Merson}, {Miller},
  {Miquel}, {Mohr}, {Morice-Atkinson}, {Naidoo}, {Neilsen}, {Nichol}, {Nord},
  {Ogando}, {Ostrovski}, {Palmese}, {Papadopoulos}, {Peiris}, {Peoples},
  {Percival}, {Plazas}, {Reed}, {Refregier}, {Romer}, {Roodman}, {Ross},
  {Rozo}, {Rykoff}, {Sadeh}, {Sako}, {S{\'a}nchez}, {Sanchez}, {Santiago},
  {Scarpine}, {Schubnell}, {Sevilla-Noarbe}, {Sheldon}, {Smith}, {Smith},
  {Soares-Santos}, {Sobreira}, {Soumagnac}, {Suchyta}, {Sullivan}, {Swanson},
  {Tarle}, {Thaler}, {Thomas}, {Thomas}, {Tucker}, {Vieira}, {Vikram},
  {Walker}, {Wechsler}, {Weller}, {Wester}, {Whiteway}, {Wilcox}, {Yanny},
  {Zhang}, \& {Zuntz}}]{DES_2016}
{Dark Energy Survey Collaboration}, {Abbott}, T., {Abdalla}, F.~B., {et~al.}
  2016, \mnras, 460, 1270

\bibitem[{{de Jong} {et~al.}(2017){de Jong}, {Kleijn}, {Erben}, {Hildebrandt},
  {Kuijken}, {Sikkema}, {Brescia}, {Bilicki}, {Napolitano}, {Amaro}, {Begeman},
  {Boxhoorn}, {Buddelmeijer}, {Cavuoti}, {Getman}, {Grado}, {Helmich}, {Huang},
  {Irisarri}, {La Barbera}, {Longo}, {McFarland}, {Nakajima}, {Paolillo},
  {Puddu}, {Radovich}, {Rifatto}, {Tortora}, {Valentijn}, {Vellucci}, {Vriend},
  {Amon}, {Blake}, {Choi}, {Conti}, {Gwyn}, {Herbonnet}, {Heymans}, {Hoekstra},
  {Klaes}, {Merten}, {Miller}, {Schneider}, \& {Viola}}]{KiDS_de_Jong_2017}
{de Jong}, J.~T.~A., {Kleijn}, G.~A.~V., {Erben}, T., {et~al.} 2017, \aap, 604,
  A134

\bibitem[{{de Jong} {et~al.}(2015){de Jong}, {Verdoes Kleijn}, {Boxhoorn},
  {Buddelmeijer}, {Capaccioli}, {Getman}, {Grado}, {Helmich}, {Huang},
  {Irisarri}, {Kuijken}, {La Barbera}, {McFarland}, {Napolitano}, {Radovich},
  {Sikkema}, {Valentijn}, {Begeman}, {Brescia}, {Cavuoti}, {Choi}, {Cordes},
  {Covone}, {Dall'Ora}, {Hildebrandt}, {Longo}, {Nakajima}, {Paolillo},
  {Puddu}, {Rifatto}, {Tortora}, {van Uitert}, {Buddendiek},
  {Harnois-D{\'e}raps}, {Erben}, {Eriksen}, {Heymans}, {Hoekstra}, {Joachimi},
  {Kitching}, {Klaes}, {Koopmans}, {K{\"o}hlinger}, {Roy}, {Sif{\'o}n},
  {Schneider}, {Sutherland}, {Viola}, \& {Vriend}}]{KiDS_de_Jong_2015}
{de Jong}, J.~T.~A., {Verdoes Kleijn}, G.~A., {Boxhoorn}, D.~R., {et~al.} 2015,
  \aap, 582, A62

\bibitem[{{de Jong} {et~al.}(2013){de Jong}, {Verdoes Kleijn}, {Kuijken}, \&
  {Valentijn}}]{KiDS_de_Jong_2013}
{de Jong}, J.~T.~A., {Verdoes Kleijn}, G.~A., {Kuijken}, K.~H., \& {Valentijn},
  E.~A. 2013, Experimental Astronomy, 35, 25

\bibitem[{{Driver} {et~al.}(2011){Driver}, {Hill}, {Kelvin}, {Robotham},
  {Liske}, {Norberg}, {Baldry}, {Bamford}, {Hopkins}, {Loveday}, {Peacock},
  {Andrae}, {Bland-Hawthorn}, {Brough}, {Brown}, {Cameron}, {Ching}, {Colless},
  {Conselice}, {Croom}, {Cross}, {de Propris}, {Dye}, {Drinkwater}, {Ellis},
  {Graham}, {Grootes}, {Gunawardhana}, {Jones}, {van Kampen}, {Maraston},
  {Nichol}, {Parkinson}, {Phillipps}, {Pimbblet}, {Popescu}, {Prescott},
  {Roseboom}, {Sadler}, {Sansom}, {Sharp}, {Smith}, {Taylor}, {Thomas},
  {Tuffs}, {Wijesinghe}, {Dunne}, {Frenk}, {Jarvis}, {Madore}, {Meyer},
  {Seibert}, {Staveley-Smith}, {Sutherland}, \& {Warren}}]{GAMA_Driver_2011}
{Driver}, S.~P., {Hill}, D.~T., {Kelvin}, L.~S., {et~al.} 2011, \mnras, 413,
  971

\bibitem[{{Duffy} {et~al.}(2008){Duffy}, {Schaye}, {Kay}, \& {Dalla
  Vecchia}}]{Halo_Duffy_2008}
{Duffy}, A.~R., {Schaye}, J., {Kay}, S.~T., \& {Dalla Vecchia}, C. 2008,
  \mnras, 390, L64

\bibitem[{{Dvornik} {et~al.}(2017){Dvornik}, {Cacciato}, {Kuijken}, {Viola},
  {Hoekstra}, {Nakajima}, {van Uitert}, {Brouwer}, {Choi}, {Erben}, {Fenech
  Conti}, {Farrow}, {Herbonnet}, {Heymans}, {Hildebrandt}, {Hopkins},
  {McFarland}, {Norberg}, {Schneider}, {Sif{\'o}n}, {Valentijn}, \&
  {Wang}}]{KiDS_Dvornik_2017}
{Dvornik}, A., {Cacciato}, M., {Kuijken}, K., {et~al.} 2017, \mnras, 468, 3251

\bibitem[{{Erben} {et~al.}(2009){Erben}, {Hildebrandt}, {Lerchster}, {Hudelot},
  {Benjamin}, {van Waerbeke}, {Schrabback}, {Brimioulle}, {Cordes}, {Dietrich},
  {Holhjem}, {Schirmer}, \& {Schneider}}]{CARS_Erben_2009}
{Erben}, T., {Hildebrandt}, H., {Lerchster}, M., {et~al.} 2009, \aap, 493, 1197

\bibitem[{{Erben} {et~al.}(2013){Erben}, {Hildebrandt}, {Miller}, {van
  Waerbeke}, {Heymans}, {Hoekstra}, {Kitching}, {Mellier}, {Benjamin}, {Blake},
  {Bonnett}, {Cordes}, {Coupon}, {Fu}, {Gavazzi}, {Gillis}, {Grocutt}, {Gwyn},
  {Holhjem}, {Hudson}, {Kilbinger}, {Kuijken}, {Milkeraitis}, {Rowe},
  {Schrabback}, {Semboloni}, {Simon}, {Smit}, {Toader}, {Vafaei}, {van Uitert},
  \& {Velander}}]{CFHTLenS_Erben_2013}
{Erben}, T., {Hildebrandt}, H., {Miller}, L., {et~al.} 2013, \mnras, 433, 2545

\bibitem[{{Erben} {et~al.}(2005){Erben}, {Schirmer}, {Dietrich}, {Cordes},
  {Haberzettl}, {Hetterscheidt}, {Hildebrandt}, {Schmithuesen}, {Schneider},
  {Simon}, {Deul}, {Hook}, {Kaiser}, {Radovich}, {Benoist}, {Nonino}, {Olsen},
  {Prandoni}, {Wichmann}, {Zaggia}, {Bomans}, {Dettmar}, \&
  {Miralles}}]{THELI_Erben_2005}
{Erben}, T., {Schirmer}, M., {Dietrich}, J.~P., {et~al.} 2005, Astronomische
  Nachrichten, 326, 432

\bibitem[{{Evans} \& {Bridle}(2009)}]{Halo_Evans_2009}
{Evans}, A. K.~D. \& {Bridle}, S. 2009, \apj, 695, 1446

\bibitem[{{Falk}(1997)}]{MAD_Falk_1997}
{Falk}, M. 1997, Annals of the Institute of Statistical Mathematics, 49, 615

\bibitem[{{Feigelson}(1988)}]{Stat_Feigelson_1988}
{Feigelson}, E.~D. 1988, Bulletin d'Information du Centre de Donnees
  Stellaires, 35, 197

\bibitem[{{Feigelson}(2009)}]{Stat_Feigelson_2009}
{Feigelson}, E.~D. 2009, arXiv e-prints, arXiv:0903.0416

\bibitem[{{Feigelson} \& {Babu}(2013)}]{Stat_Feigelson_2013}
{Feigelson}, E.~D. \& {Babu}, G.~J. 2013, {Statistical Methods for Astronomy},
  ed. T.~D. {Oswalt} \& H.~E. {Bond} (Springer, Dordrecht), 445

\bibitem[{{Fenech Conti} {et~al.}(2017){Fenech Conti}, {Herbonnet}, {Hoekstra},
  {Merten}, {Miller}, \& {Viola}}]{KiDS_Fenech_Conti_2017}
{Fenech Conti}, I., {Herbonnet}, R., {Hoekstra}, H., {et~al.} 2017, \mnras,
  467, 1627

\bibitem[{{Foreman-Mackey} {et~al.}(2013){Foreman-Mackey}, {Hogg}, {Lang}, \&
  {Goodman}}]{emcee_Foreman_2013}
{Foreman-Mackey}, D., {Hogg}, D.~W., {Lang}, D., \& {Goodman}, J. 2013, \pasp,
  125, 306

\bibitem[{{George} {et~al.}(2012){George}, {Leauthaud}, {Bundy}, {Finoguenov},
  {Ma}, {Rykoff}, {Tinker}, {Wechsler}, {Massey}, \& {Mei}}]{Halo_George_2012}
{George}, M.~R., {Leauthaud}, A., {Bundy}, K., {et~al.} 2012, \apj, 757, 2

\bibitem[{{Giocoli} {et~al.}(2021){Giocoli}, {Marulli}, {Moscardini}, {Sereno},
  {Veropalumbo}, {Gigante}, {Maturi}, {Radovich}, {Bellagamba}, {Roncarelli},
  {Bardelli}, {Contarini}, {Covone}, {Harnois-D{\'e}raps}, {Ingoglia}, {Lesci},
  {Nanni}, \& {Puddu}}]{Giocoli_2021_AMICO}
{Giocoli}, C., {Marulli}, F., {Moscardini}, L., {et~al.} 2021, \aap, 653, A19

\bibitem[{{Heck} {et~al.}(1985){Heck}, {Murtagh}, \& {Ponz}}]{Stat_Heck_1985}
{Heck}, A., {Murtagh}, F., \& {Ponz}, D. 1985, The Messenger, 41, 22

\bibitem[{{Herbonnet} {et~al.}(2017){Herbonnet}, {Buddendiek}, \&
  {Kuijken}}]{Shape_Herbonnet_2017}
{Herbonnet}, R., {Buddendiek}, A., \& {Kuijken}, K. 2017, \aap, 599, A73

\bibitem[{{Heymans} {et~al.}(2012){Heymans}, {Rowe}, {Hoekstra}, {Miller},
  {Erben}, {Kitching}, \& {van Waerbeke}}]{SYST_Heymans_2012}
{Heymans}, C., {Rowe}, B., {Hoekstra}, H., {et~al.} 2012, \mnras, 421, 381

\bibitem[{{Heymans} {et~al.}(2006){Heymans}, {Van Waerbeke}, {Bacon}, {Berge},
  {Bernstein}, {Bertin}, {Bridle}, {Brown}, {Clowe}, {Dahle}, {Erben}, {Gray},
  {Hetterscheidt}, {Hoekstra}, {Hudelot}, {Jarvis}, {Kuijken}, {Margoniner},
  {Massey}, {Mellier}, {Nakajima}, {Refregier}, {Rhodes}, {Schrabback}, \&
  {Wittman}}]{STEP_Heymans_2006}
{Heymans}, C., {Van Waerbeke}, L., {Bacon}, D., {et~al.} 2006, \mnras, 368,
  1323

\bibitem[{Heymans {et~al.}(2012)Heymans, van Waerbeke, Miller, Erben,
  Hildebrandt, Hoekstra, Kitching, Mellier, Simon, Bonnett, Coupon, Fu,
  Harnois-Déraps, Hudson, Kilbinger, Kuijken, Rowe, Schrabback, Semboloni, van
  Uitert, Vafaei, \& Velander}]{CFHTLenS_Heymans_2012}
Heymans, C., van Waerbeke, L., Miller, L., {et~al.} 2012, Monthly Notices of
  the Royal Astronomical Society, 427, 146

\bibitem[{{Hildebrandt} {et~al.}(2016){Hildebrandt}, {Choi}, {Heymans},
  {Blake}, {Erben}, {Miller}, {Nakajima}, {van Waerbeke}, {Viola},
  {Buddendiek}, {Harnois-D{\'e}raps}, {Hojjati}, {Joachimi}, {Joudaki},
  {Kitching}, {Wolf}, {Gwyn}, {Johnson}, {Kuijken}, {Sheikhbahaee}, {Tudorica},
  \& {Yee}}]{RCSLenS_2016}
{Hildebrandt}, H., {Choi}, A., {Heymans}, C., {et~al.} 2016, \mnras, 463, 635

\bibitem[{{Hildebrandt} {et~al.}(2012){Hildebrandt}, {Erben}, {Kuijken}, {van
  Waerbeke}, {Heymans}, {Coupon}, {Benjamin}, {Bonnett}, {Fu}, {Hoekstra},
  {Kitching}, {Mellier}, {Miller}, {Velander}, {Hudson}, {Rowe}, {Schrabback},
  {Semboloni}, \& {Ben{\'\i}tez}}]{CFHTLenS_Hildebrandt_2012}
{Hildebrandt}, H., {Erben}, T., {Kuijken}, K., {et~al.} 2012, \mnras, 421, 2355

\bibitem[{{Hildebrandt} {et~al.}(2017){Hildebrandt}, {Viola}, {Heymans},
  {Joudaki}, {Kuijken}, {Blake}, {Erben}, {Joachimi}, {Klaes}, {Miller},
  {Morrison}, {Nakajima}, {Verdoes Kleijn}, {Amon}, {Choi}, {Covone}, {de
  Jong}, {Dvornik}, {Fenech Conti}, {Grado}, {Harnois-D{\'e}raps}, {Herbonnet},
  {Hoekstra}, {K{\"o}hlinger}, {McFarland}, {Mead}, {Merten}, {Napolitano},
  {Peacock}, {Radovich}, {Schneider}, {Simon}, {Valentijn}, {van den Busch},
  {van Uitert}, \& {Van Waerbeke}}]{KiDS_Hildebrandt_2017}
{Hildebrandt}, H., {Viola}, M., {Heymans}, C., {et~al.} 2017, \mnras, 465, 1454

\bibitem[{{Hirata} \& {Seljak}(2003)}]{Shape_Hirata_2003}
{Hirata}, C. \& {Seljak}, U. 2003, \mnras, 343, 459

\bibitem[{{Hoekstra} \& {Jain}(2008)}]{WLgen_HJ_2008}
{Hoekstra}, H. \& {Jain}, B. 2008, Annual Review of Nuclear and Particle
  Science, 58, 99

\bibitem[{{Ivezi{\'c}} {et~al.}(2019){Ivezi{\'c}}, {Kahn}, {Tyson}, {Abel},
  {Acosta}, {Allsman}, {Alonso}, {AlSayyad}, {Anderson}, {Andrew}, {Angel},
  {Angeli}, {Ansari}, {Antilogus}, {Araujo}, {Armstrong}, {Arndt}, {Astier},
  {Aubourg}, {Auza}, {Axelrod}, {Bard}, {Barr}, {Barrau}, {Bartlett}, {Bauer},
  {Bauman}, {Baumont}, {Bechtol}, {Bechtol}, {Becker}, {Becla}, {Beldica},
  {Bellavia}, {Bianco}, {Biswas}, {Blanc}, {Blazek}, {Blandford}, {Bloom},
  {Bogart}, {Bond}, {Booth}, {Borgland}, {Borne}, {Bosch}, {Boutigny},
  {Brackett}, {Bradshaw}, {Brandt}, {Brown}, {Bullock}, {Burchat}, {Burke},
  {Cagnoli}, {Calabrese}, {Callahan}, {Callen}, {Carlin}, {Carlson},
  {Chandrasekharan}, {Charles-Emerson}, {Chesley}, {Cheu}, {Chiang}, {Chiang},
  {Chirino}, {Chow}, {Ciardi}, {Claver}, {Cohen-Tanugi}, {Cockrum}, {Coles},
  {Connolly}, {Cook}, {Cooray}, {Covey}, {Cribbs}, {Cui}, {Cutri}, {Daly},
  {Daniel}, {Daruich}, {Daubard}, {Daues}, {Dawson}, {Delgado}, {Dellapenna},
  {de Peyster}, {de Val-Borro}, {Digel}, {Doherty}, {Dubois},
  {Dubois-Felsmann}, {Durech}, {Economou}, {Eifler}, {Eracleous}, {Emmons},
  {Fausti Neto}, {Ferguson}, {Figueroa}, {Fisher-Levine}, {Focke}, {Foss},
  {Frank}, {Freemon}, {Gangler}, {Gawiser}, {Geary}, {Gee}, {Geha}, {Gessner},
  {Gibson}, {Gilmore}, {Glanzman}, {Glick}, {Goldina}, {Goldstein}, {Goodenow},
  {Graham}, {Gressler}, {Gris}, {Guy}, {Guyonnet}, {Haller}, {Harris},
  {Hascall}, {Haupt}, {Hernandez}, {Herrmann}, {Hileman}, {Hoblitt}, {Hodgson},
  {Hogan}, {Howard}, {Huang}, {Huffer}, {Ingraham}, {Innes}, {Jacoby}, {Jain},
  {Jammes}, {Jee}, {Jenness}, {Jernigan}, {Jevremovi{\'c}}, {Johns}, {Johnson},
  {Johnson}, {Jones}, {Juramy-Gilles}, {Juri{\'c}}, {Kalirai}, {Kallivayalil},
  {Kalmbach}, {Kantor}, {Karst}, {Kasliwal}, {Kelly}, {Kessler}, {Kinnison},
  {Kirkby}, {Knox}, {Kotov}, {Krabbendam}, {Krughoff}, {Kub{\'a}nek},
  {Kuczewski}, {Kulkarni}, {Ku}, {Kurita}, {Lage}, {Lambert}, {Lange},
  {Langton}, {Le Guillou}, {Levine}, {Liang}, {Lim}, {Lintott}, {Long},
  {Lopez}, {Lotz}, {Lupton}, {Lust}, {MacArthur}, {Mahabal}, {Mandelbaum},
  {Markiewicz}, {Marsh}, {Marshall}, {Marshall}, {May}, {McKercher}, {McQueen},
  {Meyers}, {Migliore}, {Miller}, {Mills}, {Miraval}, {Moeyens}, {Moolekamp},
  {Monet}, {Moniez}, {Monkewitz}, {Montgomery}, {Morrison}, {Mueller},
  {Muller}, {Mu{\~n}oz Arancibia}, {Neill}, {Newbry}, {Nief}, {Nomerotski},
  {Nordby}, {O'Connor}, {Oliver}, {Olivier}, {Olsen}, {O'Mullane}, {Ortiz},
  {Osier}, {Owen}, {Pain}, {Palecek}, {Parejko}, {Parsons}, {Pease},
  {Peterson}, {Peterson}, {Petravick}, {Libby Petrick}, {Petry},
  {Pierfederici}, {Pietrowicz}, {Pike}, {Pinto}, {Plante}, {Plate}, {Plutchak},
  {Price}, {Prouza}, {Radeka}, {Rajagopal}, {Rasmussen}, {Regnault}, {Reil},
  {Reiss}, {Reuter}, {Ridgway}, {Riot}, {Ritz}, {Robinson}, {Roby}, {Roodman},
  {Rosing}, {Roucelle}, {Rumore}, {Russo}, {Saha}, {Sassolas}, {Schalk},
  {Schellart}, {Schindler}, {Schmidt}, {Schneider}, {Schneider}, {Schoening},
  {Schumacher}, {Schwamb}, {Sebag}, {Selvy}, {Sembroski}, {Seppala}, {Serio},
  {Serrano}, {Shaw}, {Shipsey}, {Sick}, {Silvestri}, {Slater}, {Smith},
  {Smith}, {Sobhani}, {Soldahl}, {Storrie-Lombardi}, {Stover}, {Strauss},
  {Street}, {Stubbs}, {Sullivan}, {Sweeney}, {Swinbank}, {Szalay}, {Takacs},
  {Tether}, {Thaler}, {Thayer}, {Thomas}, {Thornton}, {Thukral}, {Tice},
  {Trilling}, {Turri}, {Van Berg}, {Vanden Berk}, {Vetter}, {Virieux},
  {Vucina}, {Wahl}, {Walkowicz}, {Walsh}, {Walter}, {Wang}, {Wang}, {Warner},
  {Wiecha}, {Willman}, {Winters}, {Wittman}, {Wolff}, {Wood-Vasey}, {Wu},
  {Xin}, {Yoachim}, \& {Zhan}}]{LSST_2019}
{Ivezi{\'c}}, {\v{Z}}., {Kahn}, S.~M., {Tyson}, J.~A., {et~al.} 2019, \apj,
  873, 111

\bibitem[{{Johnston} {et~al.}(2007){Johnston}, {Sheldon}, {Wechsler}, {Rozo},
  {Koester}, {Frieman}, {McKay}, {Evrard}, {Becker}, \&
  {Annis}}]{Halo_Johnston_2007}
{Johnston}, D.~E., {Sheldon}, E.~S., {Wechsler}, R.~H., {et~al.} 2007, arXiv
  e-prints, arXiv:0709.1159

\bibitem[{{Kacprzak} {et~al.}(2014){Kacprzak}, {Bridle}, {Rowe}, {Voigt},
  {Zuntz}, {Hirsch}, \& {MacCrann}}]{Bias_Kacprzak_2014}
{Kacprzak}, T., {Bridle}, S., {Rowe}, B., {et~al.} 2014, \mnras, 441, 2528

\bibitem[{{Kacprzak} {et~al.}(2012){Kacprzak}, {Zuntz}, {Rowe}, {Bridle},
  {Refregier}, {Amara}, {Voigt}, \& {Hirsch}}]{Bias_Kacprzak_2012}
{Kacprzak}, T., {Zuntz}, J., {Rowe}, B., {et~al.} 2012, \mnras, 427, 2711

\bibitem[{{Kaiser} {et~al.}(1995){Kaiser}, {Squires}, \&
  {Broadhurst}}]{Shape_Kaiser_1995}
{Kaiser}, N., {Squires}, G., \& {Broadhurst}, T. 1995, \apj, 449, 460

\bibitem[{{Kaiser} {et~al.}(2000){Kaiser}, {Wilson}, \&
  {Luppino}}]{WLhist_Kaiser_2000}
{Kaiser}, N., {Wilson}, G., \& {Luppino}, G.~A. 2000, arXiv e-prints, astro

\bibitem[{{Kautsch} {et~al.}(2008){Kautsch}, {Gonzalez}, {Soto}, {Tran},
  {Zaritsky}, \& {Moustakas}}]{Form_Kautsch_2008}
{Kautsch}, S.~J., {Gonzalez}, A.~H., {Soto}, C.~A., {et~al.} 2008, \apjl, 688,
  L5

\bibitem[{{Kitching} {et~al.}(2012){Kitching}, {Balan}, {Bridle}, {Cantale},
  {Courbin}, {Eifler}, {Gentile}, {Gill}, {Harmeling}, {Heymans}, {Hirsch},
  {Honscheid}, {Kacprzak}, {Kirkby}, {Margala}, {Massey}, {Melchior},
  {Nurbaeva}, {Patton}, {Rhodes}, {Rowe}, {Taylor}, {Tewes}, {Viola},
  {Witherick}, {Voigt}, {Young}, \& {Zuntz}}]{GREAT_Kitching_2012}
{Kitching}, T.~D., {Balan}, S.~T., {Bridle}, S., {et~al.} 2012, \mnras, 423,
  3163

\bibitem[{{Kitching} {et~al.}(2008){Kitching}, {Miller}, {Heymans}, {van
  Waerbeke}, \& {Heavens}}]{lensfit_Kitching_2008}
{Kitching}, T.~D., {Miller}, L., {Heymans}, C.~E., {van Waerbeke}, L., \&
  {Heavens}, A.~F. 2008, \mnras, 390, 149

\bibitem[{{Kuijken}(1999)}]{Shape_Kuijken_1999}
{Kuijken}, K. 1999, \aap, 352, 355

\bibitem[{{Kuijken}(2006)}]{Shape_Kuijken_2006}
{Kuijken}, K. 2006, \aap, 456, 827

\bibitem[{{Kuijken}(2011)}]{KiDS_Kuijken_2011}
{Kuijken}, K. 2011, The Messenger, 146, 8

\bibitem[{{Kuijken} {et~al.}(2015){Kuijken}, {Heymans}, {Hildebrandt},
  {Nakajima}, {Erben}, {de Jong}, {Viola}, {Choi}, {Hoekstra}, {Miller}, {van
  Uitert}, {Amon}, {Blake}, {Brouwer}, {Buddendiek}, {Conti}, {Eriksen},
  {Grado}, {Harnois-D{\'e}raps}, {Helmich}, {Herbonnet}, {Irisarri},
  {Kitching}, {Klaes}, {La Barbera}, {Napolitano}, {Radovich}, {Schneider},
  {Sif{\'o}n}, {Sikkema}, {Simon}, {Tudorica}, {Valentijn}, {Verdoes Kleijn},
  \& {van Waerbeke}}]{KiDS_Kuijken_2015}
{Kuijken}, K., {Heymans}, C., {Hildebrandt}, H., {et~al.} 2015, \mnras, 454,
  3500

\bibitem[{{Lambas} {et~al.}(1992){Lambas}, {Maddox}, \&
  {Loveday}}]{Shape_Lambas_1992}
{Lambas}, D.~G., {Maddox}, S.~J., \& {Loveday}, J. 1992, \mnras, 258, 404

\bibitem[{{Laureijs} {et~al.}(2011){Laureijs}, {Amiaux}, {Arduini},
  {Augu{\`e}res}, {Brinchmann}, {Cole}, {Cropper}, {Dabin}, {Duvet}, {Ealet},
  {Garilli}, {Gondoin}, {Guzzo}, {Hoar}, {Hoekstra}, {Holmes}, {Kitching},
  {Maciaszek}, {Mellier}, {Pasian}, {Percival}, {Rhodes}, {Saavedra Criado},
  {Sauvage}, {Scaramella}, {Valenziano}, {Warren}, {Bender}, {Castander},
  {Cimatti}, {Le F{\`e}vre}, {Kurki-Suonio}, {Levi}, {Lilje}, {Meylan},
  {Nichol}, {Pedersen}, {Popa}, {Rebolo Lopez}, {Rix}, {Rottgering},
  {Zeilinger}, {Grupp}, {Hudelot}, {Massey}, {Meneghetti}, {Miller}, {Paltani},
  {Paulin-Henriksson}, {Pires}, {Saxton}, {Schrabback}, {Seidel}, {Walsh},
  {Aghanim}, {Amendola}, {Bartlett}, {Baccigalupi}, {Beaulieu}, {Benabed},
  {Cuby}, {Elbaz}, {Fosalba}, {Gavazzi}, {Helmi}, {Hook}, {Irwin}, {Kneib},
  {Kunz}, {Mannucci}, {Moscardini}, {Tao}, {Teyssier}, {Weller}, {Zamorani},
  {Zapatero Osorio}, {Boulade}, {Foumond}, {Di Giorgio}, {Guttridge}, {James},
  {Kemp}, {Martignac}, {Spencer}, {Walton}, {Bl{\"u}mchen}, {Bonoli},
  {Bortoletto}, {Cerna}, {Corcione}, {Fabron}, {Jahnke}, {Ligori}, {Madrid},
  {Martin}, {Morgante}, {Pamplona}, {Prieto}, {Riva}, {Toledo}, {Trifoglio},
  {Zerbi}, {Abdalla}, {Douspis}, {Grenet}, {Borgani}, {Bouwens}, {Courbin},
  {Delouis}, {Dubath}, {Fontana}, {Frailis}, {Grazian}, {Koppenh{\"o}fer},
  {Mansutti}, {Melchior}, {Mignoli}, {Mohr}, {Neissner}, {Noddle}, {Poncet},
  {Scodeggio}, {Serrano}, {Shane}, {Starck}, {Surace}, {Taylor},
  {Verdoes-Kleijn}, {Vuerli}, {Williams}, {Zacchei}, {Altieri}, {Escudero
  Sanz}, {Kohley}, {Oosterbroek}, {Astier}, {Bacon}, {Bardelli}, {Baugh},
  {Bellagamba}, {Benoist}, {Bianchi}, {Biviano}, {Branchini}, {Carbone},
  {Cardone}, {Clements}, {Colombi}, {Conselice}, {Cresci}, {Deacon}, {Dunlop},
  {Fedeli}, {Fontanot}, {Franzetti}, {Giocoli}, {Garcia-Bellido}, {Gow},
  {Heavens}, {Hewett}, {Heymans}, {Holland}, {Huang}, {Ilbert}, {Joachimi},
  {Jennins}, {Kerins}, {Kiessling}, {Kirk}, {Kotak}, {Krause}, {Lahav}, {van
  Leeuwen}, {Lesgourgues}, {Lombardi}, {Magliocchetti}, {Maguire}, {Majerotto},
  {Maoli}, {Marulli}, {Maurogordato}, {McCracken}, {McLure}, {Melchiorri},
  {Merson}, {Moresco}, {Nonino}, {Norberg}, {Peacock}, {Pello}, {Penny},
  {Pettorino}, {Di Porto}, {Pozzetti}, {Quercellini}, {Radovich}, {Rassat},
  {Roche}, {Ronayette}, {Rossetti}, {Sartoris}, {Schneider}, {Semboloni},
  {Serjeant}, {Simpson}, {Skordis}, {Smadja}, {Smartt}, {Spano}, {Spiro},
  {Sullivan}, {Tilquin}, {Trotta}, {Verde}, {Wang}, {Williger}, {Zhao},
  {Zoubian}, \& {Zucca}}]{EUCLID_2011}
{Laureijs}, R., {Amiaux}, J., {Arduini}, S., {et~al.} 2011, arXiv e-prints,
  arXiv:1110.3193

\bibitem[{{Leauthaud} {et~al.}(2010){Leauthaud}, {Finoguenov}, {Kneib},
  {Taylor}, {Massey}, {Rhodes}, {Ilbert}, {Bundy}, {Tinker}, {George}, {Capak},
  {Koekemoer}, {Johnston}, {Zhang}, {Cappelluti}, {Ellis}, {Elvis}, {Giodini},
  {Heymans}, {Le F{\`e}vre}, {Lilly}, {McCracken}, {Mellier},
  {R{\'e}fr{\'e}gier}, {Salvato}, {Scoville}, {Smoot}, {Tanaka}, {Van
  Waerbeke}, \& {Wolk}}]{Form_Leauthaud_2010}
{Leauthaud}, A., {Finoguenov}, A., {Kneib}, J.-P., {et~al.} 2010, \apj, 709, 97

\bibitem[{{Leauthaud} {et~al.}(2007){Leauthaud}, {Massey}, {Kneib}, {Rhodes},
  {Johnston}, {Capak}, {Heymans}, {Ellis}, {Koekemoer}, {Le F{\`e}vre},
  {Mellier}, {R{\'e}fr{\'e}gier}, {Robin}, {Scoville}, {Tasca}, {Taylor}, \&
  {Van Waerbeke}}]{COSMOS_2007}
{Leauthaud}, A., {Massey}, R., {Kneib}, J.-P., {et~al.} 2007, \apjs, 172, 219

\bibitem[{{Lesci} {et~al.}(in prep.){Lesci}, {Marulli}, {Moscardini}, {Sereno},
  {Veropalumb0}, {Maturi}, {Radovich}, {Bellagamba}, {Roncarelli}, {Giocoli},
  {Contarini}, {Nanni}, {Bardelli}, {Puddu}, {Covone}, \&
  {Ingoglia}}]{AMICO_Lesci_2020}
{Lesci}, G., {Marulli}, F., {Moscardini}, L., {et~al.} in prep., \aap

\bibitem[{{Mandelbaum}(2018)}]{Bias_Mandelbaum_2018}
{Mandelbaum}, R. 2018, \araa, 56, 393

\bibitem[{{Mandelbaum} {et~al.}(2015){Mandelbaum}, {Rowe}, {Armstrong}, {Bard},
  {Bertin}, {Bosch}, {Boutigny}, {Courbin}, {Dawson}, {Donnarumma}, {Fenech
  Conti}, {Gavazzi}, {Gentile}, {Gill}, {Hogg}, {Huff}, {Jee}, {Kacprzak},
  {Kilbinger}, {Kuntzer}, {Lang}, {Luo}, {March}, {Marshall}, {Meyers},
  {Miller}, {Miyatake}, {Nakajima}, {Ngol{\'e} Mboula}, {Nurbaeva}, {Okura},
  {Paulin-Henriksson}, {Rhodes}, {Schneider}, {Shan}, {Sheldon}, {Simet},
  {Starck}, {Sureau}, {Tewes}, {Zarb Adami}, {Zhang}, \&
  {Zuntz}}]{GREAT_Mandelbaum_2015}
{Mandelbaum}, R., {Rowe}, B., {Armstrong}, R., {et~al.} 2015, \mnras, 450, 2963

\bibitem[{{Massey} {et~al.}(2007){Massey}, {Heymans}, {Berg{\'e}}, {Bernstein},
  {Bridle}, {Clowe}, {Dahle}, {Ellis}, {Erben}, {Hetterscheidt}, {High},
  {Hirata}, {Hoekstra}, {Hudelot}, {Jarvis}, {Johnston}, {Kuijken},
  {Margoniner}, {Mandelbaum}, {Mellier}, {Nakajima}, {Paulin-Henriksson},
  {Peeples}, {Roat}, {Refregier}, {Rhodes}, {Schrabback}, {Schirmer}, {Seljak},
  {Semboloni}, \& {van Waerbeke}}]{STEP_Massey_2007}
{Massey}, R., {Heymans}, C., {Berg{\'e}}, J., {et~al.} 2007, \mnras, 376, 13

\bibitem[{{Maturi} {et~al.}(2019){Maturi}, {Bellagamba}, {Radovich},
  {Roncarelli}, {Sereno}, {Moscardini}, {Bardelli}, \&
  {Puddu}}]{AMICO_Maturi_2019}
{Maturi}, M., {Bellagamba}, F., {Radovich}, M., {et~al.} 2019, \mnras, 485, 498

\bibitem[{{McFarland} {et~al.}(2013){McFarland}, {Verdoes-Kleijn}, {Sikkema},
  {Helmich}, {Boxhoorn}, \& {Valentijn}}]{AWE_McFarland_2013}
{McFarland}, J.~P., {Verdoes-Kleijn}, G., {Sikkema}, G., {et~al.} 2013,
  Experimental Astronomy, 35, 45

\bibitem[{{Mead} {et~al.}(2015){Mead}, {Peacock}, {Heymans}, {Joudaki}, \&
  {Heavens}}]{Halo_Mead_2015}
{Mead}, A.~J., {Peacock}, J.~A., {Heymans}, C., {Joudaki}, S., \& {Heavens},
  A.~F. 2015, \mnras, 454, 1958

\bibitem[{{Melchior} \& {Viola}(2012)}]{Bias_Melchior_2012}
{Melchior}, P. \& {Viola}, M. 2012, \mnras, 424, 2757

\bibitem[{{Miller} {et~al.}(2013){Miller}, {Heymans}, {Kitching}, {van
  Waerbeke}, {Erben}, {Hildebrandt}, {Hoekstra}, {Mellier}, {Rowe}, {Coupon},
  {Dietrich}, {Fu}, {Harnois-D{\'e}raps}, {Hudson}, {Kilbinger}, {Kuijken},
  {Schrabback}, {Semboloni}, {Vafaei}, \& {Veland er}}]{lensfit_Miller_2013}
{Miller}, L., {Heymans}, C., {Kitching}, T.~D., {et~al.} 2013, \mnras, 429,
  2858

\bibitem[{{Miller} {et~al.}(2007){Miller}, {Kitching}, {Heymans}, {Heavens}, \&
  {van Waerbeke}}]{lensfit_Miller_2007}
{Miller}, L., {Kitching}, T.~D., {Heymans}, C., {Heavens}, A.~F., \& {van
  Waerbeke}, L. 2007, \mnras, 382, 315

\bibitem[{{Miyatake} {et~al.}(2015){Miyatake}, {More}, {Mandelbaum}, {Takada},
  {Spergel}, {Kneib}, {Schneider}, {Brinkmann}, \&
  {Brownstein}}]{Miyatake_2015}
{Miyatake}, H., {More}, S., {Mandelbaum}, R., {et~al.} 2015, \apj, 806, 1

\bibitem[{{Navarro} {et~al.}(1995){Navarro}, {Frenk}, \& {White}}]{NFW}
{Navarro}, J.~F., {Frenk}, C.~S., \& {White}, S. D.~M. 1995, \mnras, 275, 720

\bibitem[{{Oguri} {et~al.}(2012){Oguri}, {Bayliss}, {Dahle}, {Sharon},
  {Gladders}, {Natarajan}, {Hennawi}, \& {Koester}}]{Oguri_2012}
{Oguri}, M., {Bayliss}, M.~B., {Dahle}, H., {et~al.} 2012, \mnras, 420, 3213

\bibitem[{{Oguri} {et~al.}(2018){Oguri}, {Miyazaki}, {Hikage}, {Mandelbaum},
  {Utsumi}, {Miyatake}, {Takada}, {Armstrong}, {Bosch}, {Komiyama},
  {Leauthaud}, {More}, {Nishizawa}, {Okabe}, \& {Tanaka}}]{Halo_Oguri_2018}
{Oguri}, M., {Miyazaki}, S., {Hikage}, C., {et~al.} 2018, \pasj, 70, S26

\bibitem[{{Oguri} {et~al.}(2010){Oguri}, {Takada}, {Okabe}, \&
  {Smith}}]{Halo_Oguri_2010}
{Oguri}, M., {Takada}, M., {Okabe}, N., \& {Smith}, G.~P. 2010, \mnras, 405,
  2215

\bibitem[{{Peacock} \& {Smith}(2000)}]{Halo_Peacock_2000}
{Peacock}, J.~A. \& {Smith}, R.~E. 2000, \mnras, 318, 1144

\bibitem[{{Planck Collaboration} {et~al.}(2014){Planck Collaboration}, {Ade},
  {Aghanim}, {Armitage-Caplan}, {Arnaud}, {Ashdown}, {Atrio-Barandela},
  {Aumont}, {Baccigalupi}, {Banday}, \& et~al.}]{Planck_XVI_2014}
{Planck Collaboration}, {Ade}, P.~A.~R., {Aghanim}, N., {et~al.} 2014, \aap,
  571, A16

\bibitem[{{Radovich} {et~al.}(2017){Radovich}, {Puddu}, {Bellagamba},
  {Roncarelli}, {Moscardini}, {Bardelli}, {Grado}, {Getman}, {Maturi}, {Huang},
  {Napolitano}, {McFarland}, {Valentijn}, \& {Bilicki}}]{AMICO_Radovich_2017}
{Radovich}, M., {Puddu}, E., {Bellagamba}, F., {et~al.} 2017, \aap, 598, A107

\bibitem[{Rao(1945)}]{Rao_1945}
Rao, C.~R. 1945, Bull. Calcutta Math. Soc., 37, 81

\bibitem[{{Refregier} \& {Bacon}(2003)}]{Shape_Refregier_2003}
{Refregier}, A. \& {Bacon}, D. 2003, \mnras, 338, 48

\bibitem[{{Refregier} {et~al.}(2012){Refregier}, {Kacprzak}, {Amara}, {Bridle},
  \& {Rowe}}]{Bias_Refregier_2012}
{Refregier}, A., {Kacprzak}, T., {Amara}, A., {Bridle}, S., \& {Rowe}, B. 2012,
  \mnras, 425, 1951

\bibitem[{{Rhodes} {et~al.}(2000){Rhodes}, {Refregier}, \&
  {Groth}}]{Shape_Rhodes_2000}
{Rhodes}, J., {Refregier}, A., \& {Groth}, E.~J. 2000, \apj, 536, 79

\bibitem[{{Robotham} {et~al.}(2011){Robotham}, {Norberg}, {Driver}, {Baldry},
  {Bamford}, {Hopkins}, {Liske}, {Loveday}, {Merson}, {Peacock}, {Brough},
  {Cameron}, {Conselice}, {Croom}, {Frenk}, {Gunawardhana}, {Hill}, {Jones},
  {Kelvin}, {Kuijken}, {Nichol}, {Parkinson}, {Pimbblet}, {Phillipps},
  {Popescu}, {Prescott}, {Sharp}, {Sutherland}, {Taylor}, {Thomas}, {Tuffs},
  {van Kampen}, \& {Wijesinghe}}]{GAMA_Robotham_2011}
{Robotham}, A.~S.~G., {Norberg}, P., {Driver}, S.~P., {et~al.} 2011, \mnras,
  416, 2640

\bibitem[{{Rodr{\'\i}guez} \& {Padilla}(2013)}]{Shape_Rodriguez_2013}
{Rodr{\'\i}guez}, S. \& {Padilla}, N.~D. 2013, \mnras, 434, 2153

\bibitem[{{Rykoff} {et~al.}(2016){Rykoff}, {Rozo}, {Hollowood},
  {Bermeo-Hernandez}, {Jeltema}, {Mayers}, {Romer}, {Rooney}, {Saro}, {Vergara
  Cervantes}, {Wechsler}, {Wilcox}, {Abbott}, {Abdalla}, {Allam}, {Annis},
  {Benoit-L{\'e}vy}, {Bernstein}, {Bertin}, {Brooks}, {Burke}, {Capozzi},
  {Carnero Rosell}, {Carrasco Kind}, {Castander}, {Childress}, {Collins},
  {Cunha}, {D'Andrea}, {da Costa}, {Davis}, {Desai}, {Diehl}, {Dietrich},
  {Doel}, {Evrard}, {Finley}, {Flaugher}, {Fosalba}, {Frieman}, {Glazebrook},
  {Goldstein}, {Gruen}, {Gruendl}, {Gutierrez}, {Hilton}, {Honscheid}, {Hoyle},
  {James}, {Kay}, {Kuehn}, {Kuropatkin}, {Lahav}, {Lewis}, {Lidman}, {Lima},
  {Maia}, {Mann}, {Marshall}, {Martini}, {Melchior}, {Miller}, {Miquel},
  {Mohr}, {Nichol}, {Nord}, {Ogando}, {Plazas}, {Reil}, {Sahl{\'e}n},
  {Sanchez}, {Santiago}, {Scarpine}, {Schubnell}, {Sevilla-Noarbe}, {Smith},
  {Soares-Santos}, {Sobreira}, {Stott}, {Suchyta}, {Swanson}, {Tarle},
  {Thomas}, {Tucker}, {Uddin}, {Viana}, {Vikram}, {Walker}, {Zhang}, \& {DES
  Collaboration}}]{Halo_Rykoff_2016}
{Rykoff}, E.~S., {Rozo}, E., {Hollowood}, D., {et~al.} 2016, \apjs, 224, 1

\bibitem[{{Schirmer}(2013)}]{THELI_Schirmer_2013}
{Schirmer}, M. 2013, \apjs, 209, 21

\bibitem[{{Schneider} {et~al.}(2015){Schneider}, {Hogg}, {Marshall}, {Dawson},
  {Meyers}, {Bard}, \& {Lang}}]{Shape_Schneider_2015}
{Schneider}, M.~D., {Hogg}, D.~W., {Marshall}, P.~J., {et~al.} 2015, \apj, 807,
  87

\bibitem[{{Schneider}(2003)}]{WLgen_S_2003}
{Schneider}, P. 2003, arXiv e-prints, arXiv:0306465

\bibitem[{{Schneider}(2006)}]{WLgen_S_2006}
{Schneider}, P. 2006, in Saas-Fee Advanced Course 33: Gravitational Lensing:
  Strong, Weak and Micro, ed. G.~{Meylan}, P.~{Jetzer}, P.~{North},
  P.~{Schneider}, C.~S. {Kochanek}, \& J.~{Wambsganss}, 269--451

\bibitem[{{Seitz} \& {Schneider}(1995)}]{WLgen_SS_1995}
{Seitz}, C. \& {Schneider}, P. 1995, \aap, 297, 287

\bibitem[{{Seitz} \& {Schneider}(1997)}]{WLgen_SS_1997}
{Seitz}, C. \& {Schneider}, P. 1997, \aap, 318, 687

\bibitem[{{Seljak}(2000)}]{Halo_Seljak_2000}
{Seljak}, U. 2000, \mnras, 318, 203

\bibitem[{{Sellentin} {et~al.}(2018){Sellentin}, {Heymans}, \&
  {Harnois-D{\'e}raps}}]{Bias_Sellentin_2018}
{Sellentin}, E., {Heymans}, C., \& {Harnois-D{\'e}raps}, J. 2018, \mnras, 477,
  4879

\bibitem[{{Sif{\'o}n} {et~al.}(2015){Sif{\'o}n}, {Cacciato}, {Hoekstra},
  {Brouwer}, {van Uitert}, {Viola}, {Baldry}, {Brough}, {Brown}, {Choi},
  {Driver}, {Erben}, {Grado}, {Heymans}, {Hildebrandt}, {Joachimi}, {de Jong},
  {Kuijken}, {McFarland}, {Miller}, {Nakajima}, {Napolitano}, {Norberg},
  {Robotham}, {Schneider}, \& {Verdoes Kleijn}}]{KiDS_Sifon_2015}
{Sif{\'o}n}, C., {Cacciato}, M., {Hoekstra}, H., {et~al.} 2015, \mnras, 454,
  3938

\bibitem[{{Smit} \& {Kuijken}(2018)}]{PaperI}
{Smit}, M. \& {Kuijken}, K. 2018, \aap, 609, A103

\bibitem[{{Tinker} {et~al.}(2010){Tinker}, {Robertson}, {Kravtsov}, {Klypin},
  {Warren}, {Yepes}, \& {Gottl{\"o}ber}}]{Halo_Tinker_2010}
{Tinker}, J.~L., {Robertson}, B.~E., {Kravtsov}, A.~V., {et~al.} 2010, \apj,
  724, 878

\bibitem[{{Tyson} {et~al.}(1990){Tyson}, {Valdes}, \&
  {Wenk}}]{WLhist_Tyson_1990}
{Tyson}, J.~A., {Valdes}, F., \& {Wenk}, R.~A. 1990, \apjl, 349, L1

\bibitem[{{Valentijn} {et~al.}(2007){Valentijn}, {McFarland}, {Snigula},
  {Begeman}, {Boxhoorn}, {Rengelink}, {Helmich}, {Heraudeau}, {Verdoes Kleijn},
  {Vermeij}, {Vriend}, {Tempelaar}, {Deul}, {Kuijken}, {Capaccioli},
  {Silvotti}, {Bender}, {Neeser}, {Saglia}, {Bertin}, \&
  {Mellier}}]{AWE_Valentijn_2007}
{Valentijn}, E.~A., {McFarland}, J.~P., {Snigula}, J., {et~al.} 2007, in
  Astronomical Society of the Pacific Conference Series, Vol. 376, Astronomical
  Data Analysis Software and Systems XVI, ed. R.~A. {Shaw}, F.~{Hill}, \& D.~J.
  {Bell}, 491

\bibitem[{{van den Bosch} {et~al.}(2013){van den Bosch}, {More}, {Cacciato},
  {Mo}, \& {Yang}}]{Halo_Bosch_2013}
{van den Bosch}, F.~C., {More}, S., {Cacciato}, M., {Mo}, H., \& {Yang}, X.
  2013, \mnras, 430, 725

\bibitem[{{van Uitert} {et~al.}(2016){van Uitert}, {Cacciato}, {Hoekstra},
  {Brouwer}, {Sif{\'o}n}, {Viola}, {Baldry}, {Bland-Hawthorn}, {Brough},
  {Brown}, {Choi}, {Driver}, {Erben}, {Heymans}, {Hildebrandt}, {Joachimi},
  {Kuijken}, {Liske}, {Loveday}, {McFarland}, {Miller}, {Nakajima}, {Peacock},
  {Radovich}, {Robotham}, {Schneider}, {Sikkema}, {Taylor}, \& {Verdoes
  Kleijn}}]{KiDS_Uitert_2016}
{van Uitert}, E., {Cacciato}, M., {Hoekstra}, H., {et~al.} 2016, \mnras, 459,
  3251

\bibitem[{{van Uitert} {et~al.}(2017){van Uitert}, {Hoekstra}, {Joachimi},
  {Schneider}, {Bland-Hawthorn}, {Choi}, {Erben}, {Heymans}, {Hildebrandt},
  {Hopkins}, {Klaes}, {Kuijken}, {Nakajima}, {Napolitano}, {Schrabback},
  {Valentijn}, \& {Viola}}]{KiDS_Uitert_2017}
{van Uitert}, E., {Hoekstra}, H., {Joachimi}, B., {et~al.} 2017, \mnras, 467,
  4131

\bibitem[{{Van Waerbeke} {et~al.}(2000){Van Waerbeke}, {Mellier}, {Erben},
  {Cuillandre}, {Bernardeau}, {Maoli}, {Bertin}, {McCracken}, {Le F{\`e}vre},
  {Fort}, {Dantel-Fort}, {Jain}, \& {Schneider}}]{WLhist_Waerbeke_2000}
{Van Waerbeke}, L., {Mellier}, Y., {Erben}, T., {et~al.} 2000, \aap, 358, 30

\bibitem[{{Verdoes Kleijn} {et~al.}(2012){Verdoes Kleijn}, {de Jong},
  {Valentijn}, {Kuijken}, {KiDS Consortium}, \& {Astro-WISE
  Consortium}}]{AWE_Verdoes_2012}
{Verdoes Kleijn}, G., {de Jong}, J.~T.~A., {Valentijn}, E., {et~al.} 2012, in
  Astronomical Society of the Pacific Conference Series, Vol. 461, Astronomical
  Data Analysis Software and Systems XXI, ed. P.~{Ballester}, D.~{Egret}, \&
  N.~P.~F. {Lorente}, 237

\bibitem[{{Viola} {et~al.}(2015){Viola}, {Cacciato}, {Brouwer}, {Kuijken},
  {Hoekstra}, {Norberg}, {Robotham}, {van Uitert}, {Alpaslan}, {Baldry},
  {Choi}, {de Jong}, {Driver}, {Erben}, {Grado}, {Graham}, {Heymans},
  {Hildebrandt}, {Hopkins}, {Irisarri}, {Joachimi}, {Loveday}, {Miller},
  {Nakajima}, {Schneider}, {Sif{\'o}n}, \& {Verdoes Kleijn}}]{KiDS_Viola_2015}
{Viola}, M., {Cacciato}, M., {Brouwer}, M., {et~al.} 2015, \mnras, 452, 3529

\bibitem[{{Viola} {et~al.}(2014){Viola}, {Kitching}, \&
  {Joachimi}}]{Bias_Viola_2014}
{Viola}, M., {Kitching}, T.~D., \& {Joachimi}, B. 2014, \mnras, 439, 1909

\bibitem[{{Voigt} \& {Bridle}(2010)}]{SYST_Voigt_2010}
{Voigt}, L.~M. \& {Bridle}, S.~L. 2010, \mnras, 404, 458

\bibitem[{{Wittman} {et~al.}(2000){Wittman}, {Tyson}, {Kirkman},
  {Dell'Antonio}, \& {Bernstein}}]{WLhist_Wittman_2000}
{Wittman}, D.~M., {Tyson}, J.~A., {Kirkman}, D., {Dell'Antonio}, I., \&
  {Bernstein}, G. 2000, \nat, 405, 143

\end{thebibliography}

\appendix

\section{Tests for systematics} \label{app:syst}

For completeness, we repeated some of the tests for systematics that were already carried out in \citet{KiDS_Dvornik_2017} and \citet{AMICO_Bellagamba_2019} because of our difference in sky coverage, background selection, and estimated redshift distribution compared with those two studies.
% \LEt{ Verify that your intended meaning has not been changed.}

\subsection{Photometric redshift} \label{app:photoz}

%\begin{itemize}
%  \item{Estimated sources before clusters $2\%$, but corrected for by %$w_{\mathrm{ls}}=0$}
%  \item{COL selected sources also estimated with DIR method}
%  \item{Estimated confidence in $\Sigma_{\mathrm{cr}}$ from DIR method}
%  \item{Result: uncertainty from photo-$z$ is negligible ($4.2\%$ in %Bellagamba)}
%\end{itemize}

We used the same method as \citet{KiDS_Dvornik_2017} to determine the comoving critical density. There are two important differences that could affect the uncertainty in $\Sigma_{\mathrm{cr}}$: We selected lenses at a significantly higher redshift, and we complemented our background source selection with the color selection described in Sect. \ref{sec:sources}.

\begin{figure}[h]
\centering
\resizebox{\hsize}{!}{
\includegraphics[angle=90]{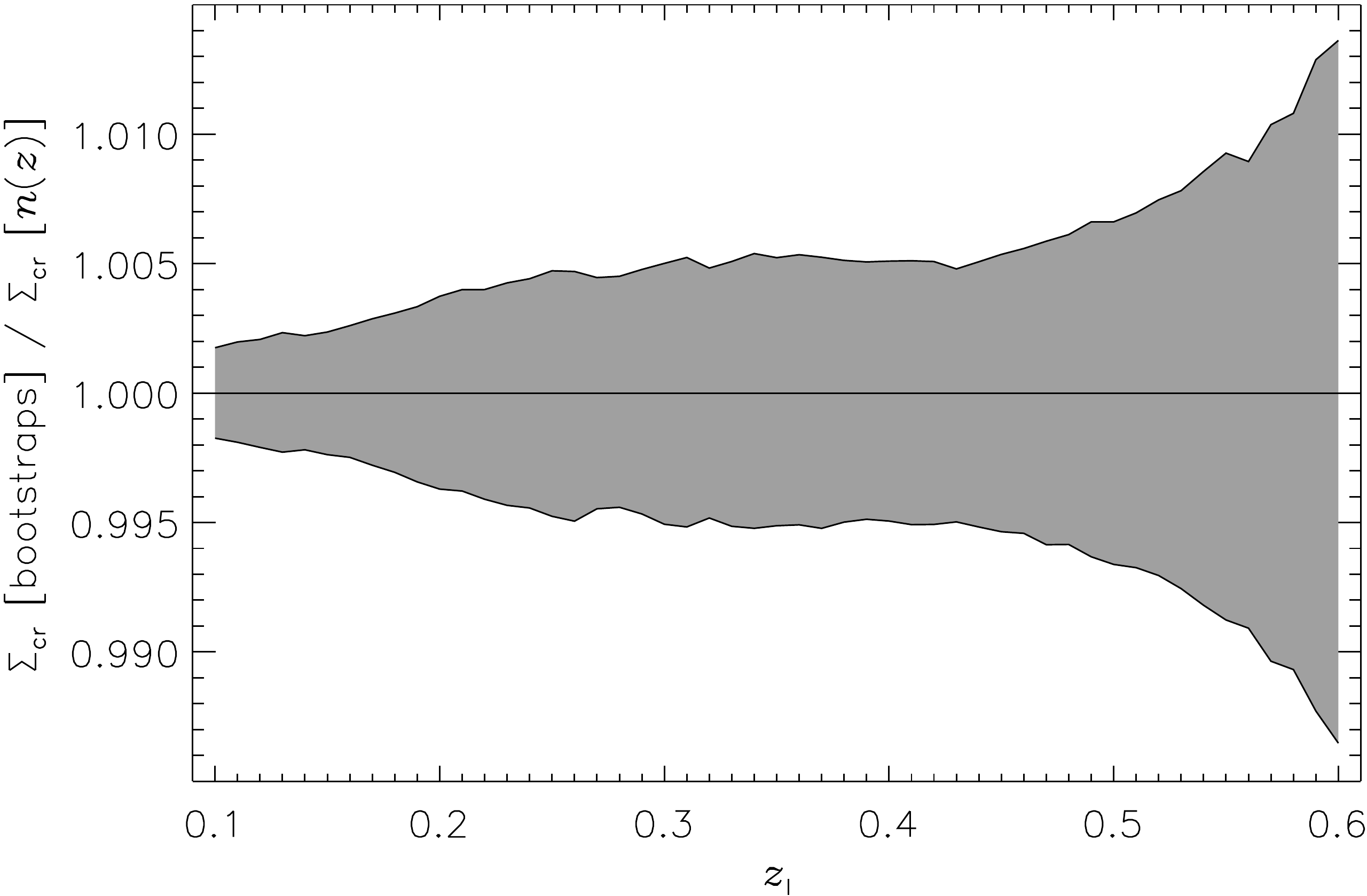}
}
\caption{Relative errors in $\Sigma_{\mathrm{cr}}$, estimated using $10^4$ bootstraps of the spectroscopic catalog of \citet{KiDS_Hildebrandt_2017}. The relative error on the ESD is negligible.}
\label{fig:conf_nz}
\end{figure}

We assessed the relative errors in $\Sigma_{\mathrm{cr}}$ by performing $10^4$ bootstraps of the spectroscopic catalog of \cite{KiDS_Hildebrandt_2017}. We find the median error on $\Sigma_{\mathrm{cr}}$ to be $\sim 0.5\%$, as shown in Fig. \ref{fig:conf_nz}.

\subsection{Contamination of the background sample by cluster galaxies} \label{app:clus_bg}

\citet{KiDS_Dvornik_2017} showed that an offset of $\Delta z = 0.2$ is enough to avoid a significant contamination of the background sources by unidentified GAMA group members. For lenses at a higher redshift, this contamination increases, while at the same time the density of available background sources decreases due to the observed depth of KiDS-450.

\begin{figure}[h]
\centering
\resizebox{\hsize}{!}{
\includegraphics[angle=90]{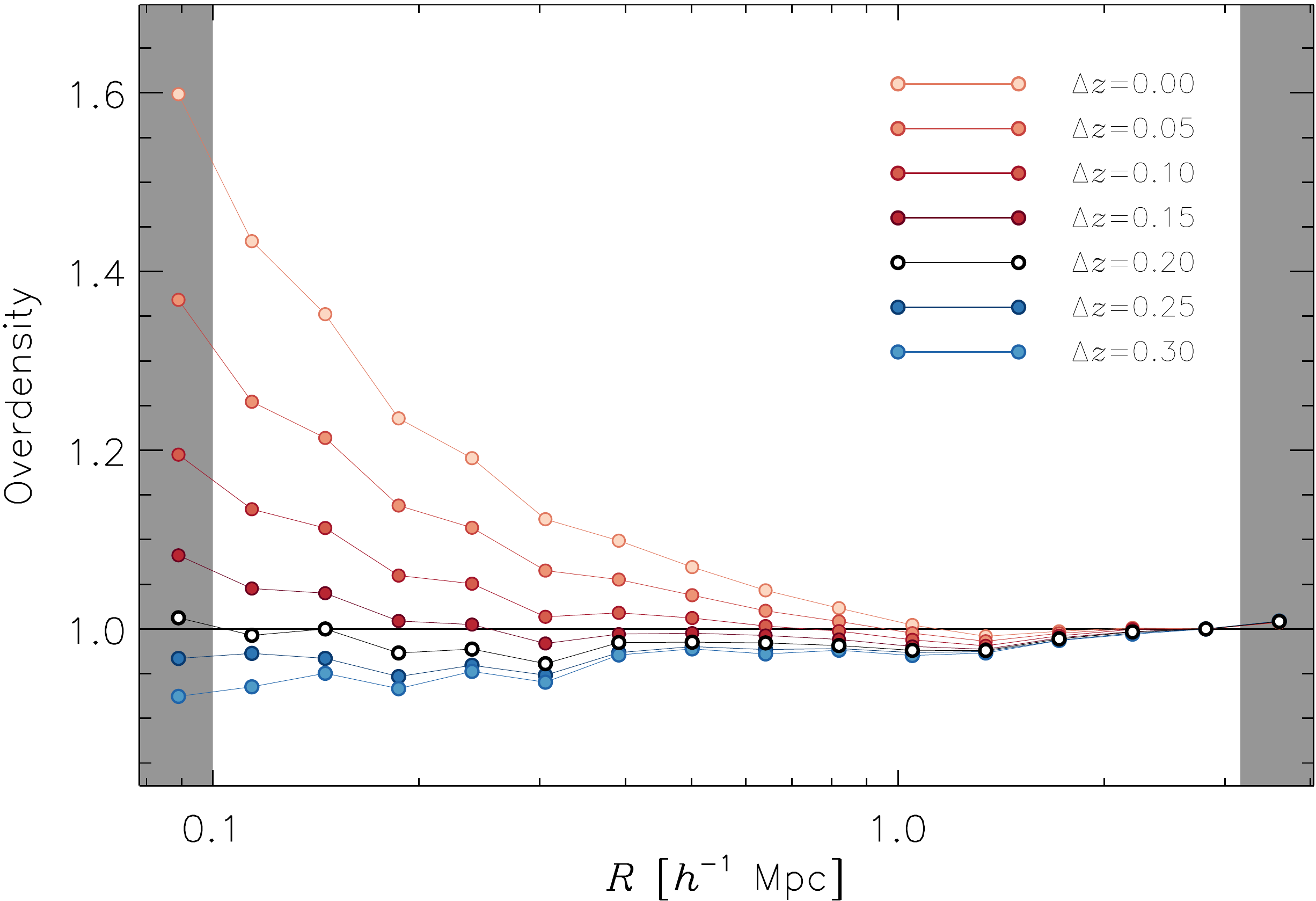}
}
\caption{Relative source densities around AMICO clusters as a function of radius, $R$, and the photometric redshift offset, $z_{\mathrm{B}}\ge z_{\mathrm{l}}+\Delta z$, between the lens and the source. We note that some small under-densities around $R=0.2$ $h^{-1}$ Mpc may be due to the relative normalization.}
\label{fig:clus_bg}
\end{figure}

We used the same test as \citet{KiDS_Dvornik_2017} to assess the source density around AMICO clusters in order to determine the necessary $\Delta z$ offset between the lens and the sources. We find that $\Delta z = 0.2$ is appropriate for our cluster selection (Fig. \ref{fig:clus_bg}).

\subsection{Individual bin ESD profiles and cross signals} \label{app:cross}

\begin{figure*}[h]
\centering
\resizebox{\hsize}{!}{
\includegraphics[angle=90]{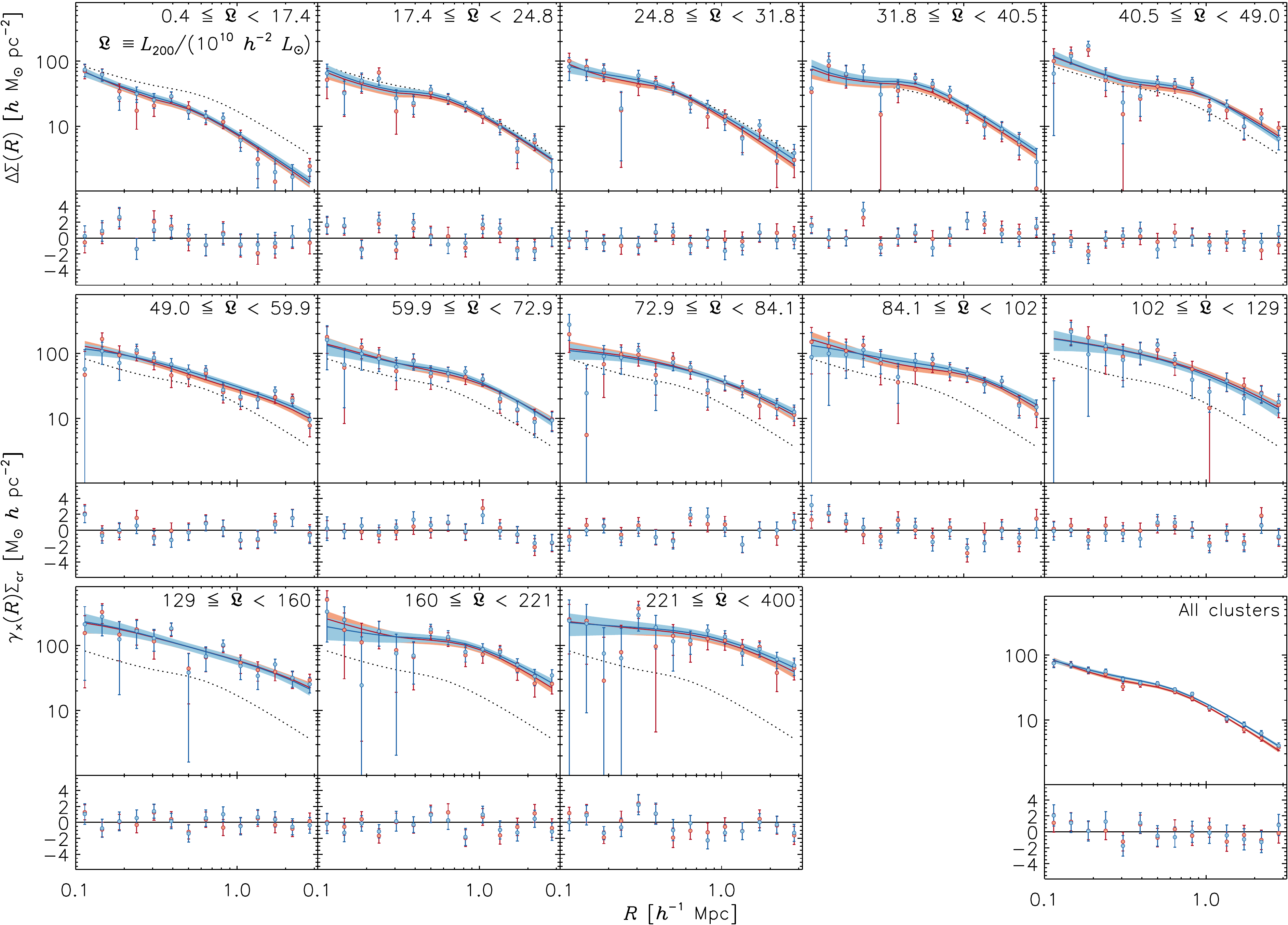}
}
\caption{Estimated ESD profiles and cross signal of the 13 cluster bins, using the LAD estimator (blue) and a weighted mean (red). The error bars are the square roots of the diagonal values of the respective covariance matrices. The solid lines represent the best fitting halo model obtained by the MCMC fit. The shaded regions show the $68.3\%$ confidence bands, estimated using the $15.9$th and $84.1$th percentiles of the MCMC realizations. Individual bins are indicated by the range in normalized luminosity, $\mathfrak{L}\equiv L_{200}\, / ( 10^{10} h^{-2} L_{\odot} )$. The lower-right plot shows the ESD profile estimated from the full AMICO cluster catalog, also shown in Fig. \ref{fig:ESD}. The average of the two best fitting halo models from Fig. \ref{fig:ESD} are shown in each panel as a dotted line for easy comparison.}
\label{fig:cross}
\end{figure*}

In Fig. \ref{fig:cross} we show the ESD profiles for the 13 cluster bins, including the cross signal, which is consistent with zero. We also show the derived halo model fits and their confidence intervals, comparing the fits using the full AMICO cluster catalog from Fig. \ref{fig:ESD}.

% BB: 13 -1050 792 35 (ps)
% BB: 13 0 792 1085 (eps)

\subsection{Tile bootstrap} \label{app:tile_bs}

\begin{figure*}[h]
\centering
\resizebox{\hsize}{!}{
\includegraphics{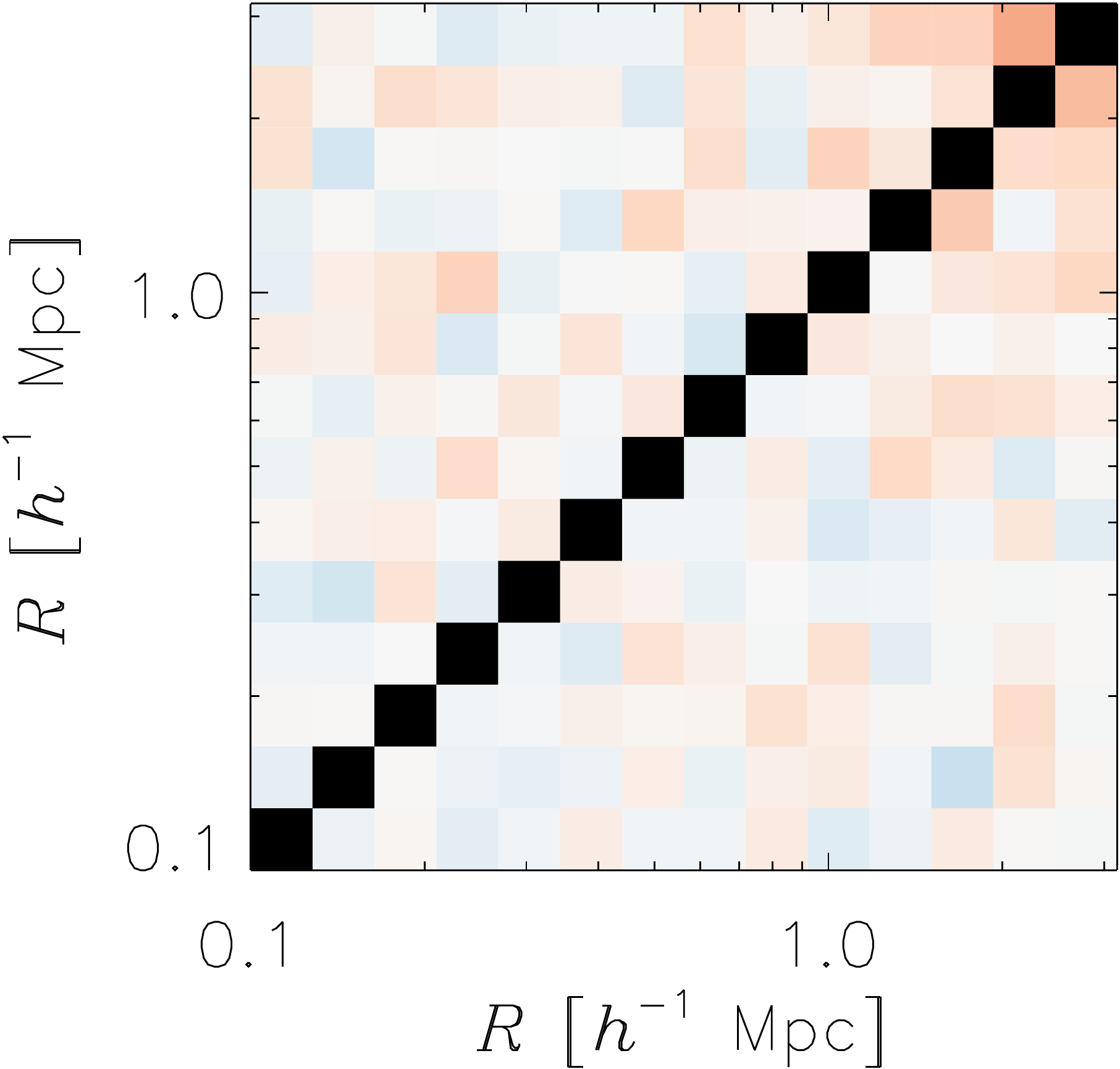}
\includegraphics{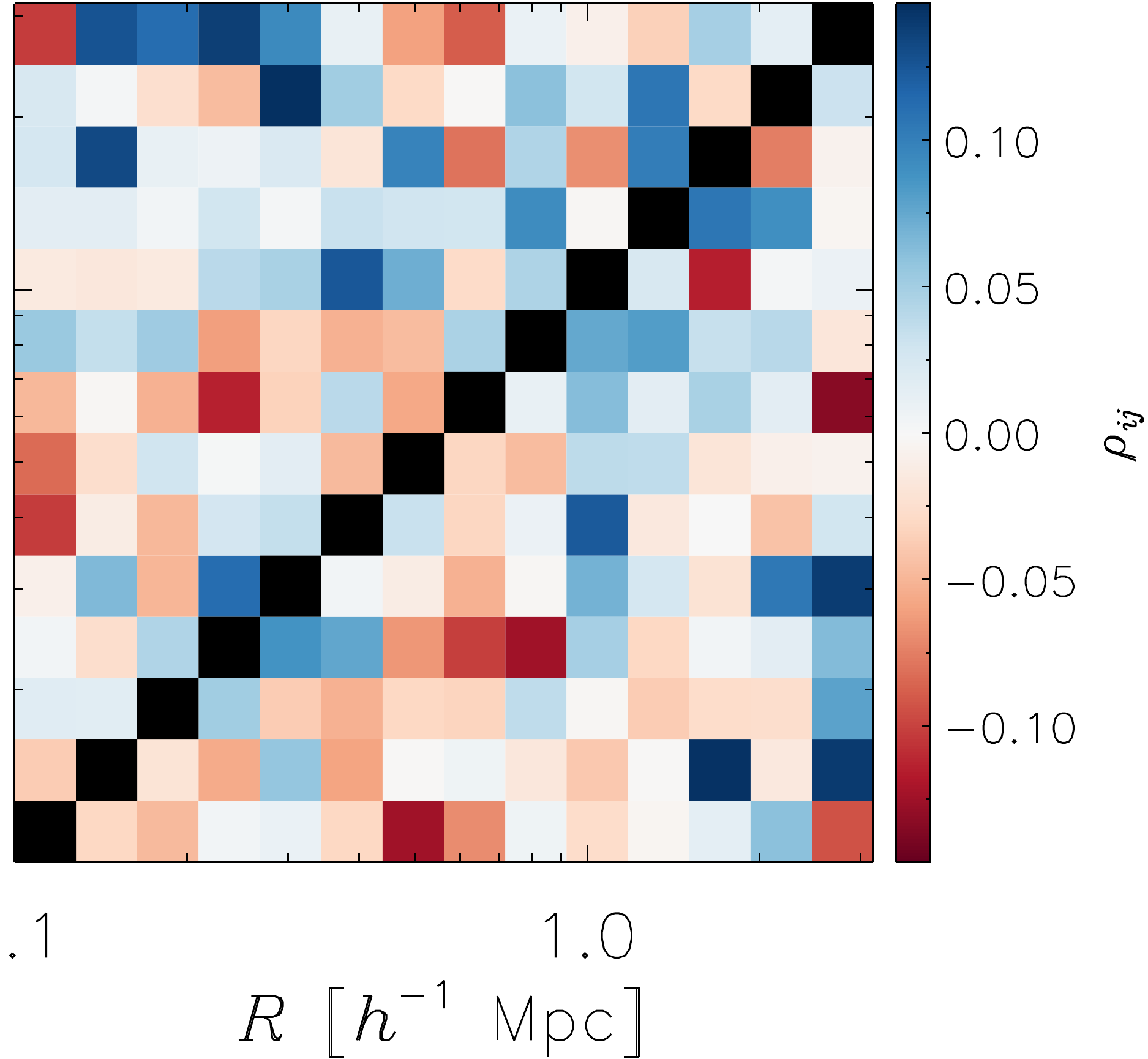}
\includegraphics{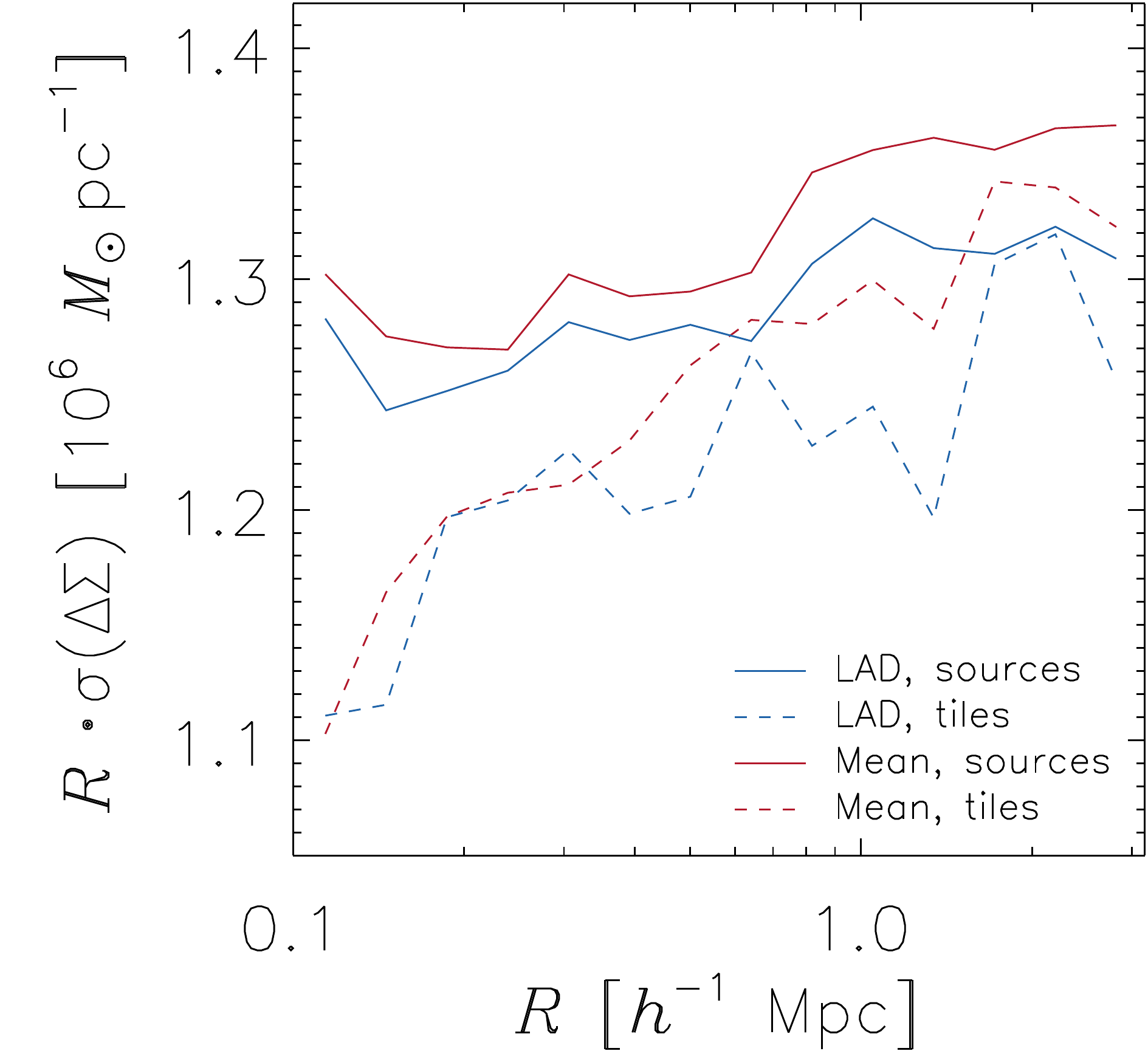}
}
\caption{Left: Correlation derived from bootstrapping the signal by individual sources; same as Fig. \ref{fig:COV}, but with the color stretch adjusted to the middle plot. Middle: Correlation derived from bootstrapping the signal in 1 deg$^2$ tiles. The upper-left corners show the correlations from LAD regression. The lower-right corners show the correlations from using the weighted mean. Right: Comparison of the errors obtained from bootstrapping sources (LAD: solid blue; mean: solid red) and bootstrapping 1 deg$^2$ tiles (LAD: dashed blue; mean: dashed red).}
\label{fig:tile_bs}
\end{figure*}

As described in Sect. \ref{sec:imp}, we could not use the same bootstrap approach as \citet{KiDS_Dvornik_2017}, due to the sparsity of lenses in the highest lens luminosity bins. Since our bootstrap approach described in Sect. \ref{sec:imp} does not account for cosmic variance and is not sensitive to the clustering effect of dark matter halos, we compare the covariances derived by the two bootstrap methods for the ESD of the whole lens selection in Fig. \ref{fig:tile_bs} and find no significant differences or pattern beyond what is expected from statistical noise. Since we expect the contribution from cosmic variance to be even lower for subsets of lenses, we conclude that our bootstrap approach yields a good estimate of the covariance. 

% BB source: 0 15 522 513
% BB tile:  75 15 615 513
% BB diag: 0 15 544 513

\section{Analysis of dependence on outer data points} \label{app:data_consistency} 

In Fig. \ref{fig:powerlaw} it can be seen that the distribution of clusters in the two outermost luminosity bins is not symmetric. At the lower end, this is due to the selection criterion of $\lambda^*$ in the AMICO catalog. At the higher end, we have only a few clusters.

We assessed the effect these two points have on the $L_{200}-M_{200}$ scaling relation by repeating the fit without these bins. We find no difference within the statistical uncertainties, as given in Eq. \ref{eq:outer_bins}:

\begin{subequations}
  \label{eq:outer_bins}
  \begin{align}
    \mathrm{Mean} \quad & \frac{M_{200}}{10^{14.1} h^{-1} M_{\odot}} = (0.98 \pm 0.06) \left( \frac{L_{200}}{10^{11.8} h^{-2} L_{\odot}} \right)^{(1.25 \pm 0.10)} \,, \\
    \mathrm{LAD} \quad & \frac{M_{200}}{10^{14.1} h^{-1} M_{\odot}} = (1.02 \pm 0.06) \left( \frac{L_{200}}{10^{11.8} h^{-2} L_{\odot}} \right)^{(1.23 \pm 0.09)} .
    \end{align}
\end{subequations}

\end{document}